# Bidirectional Vectorial Holography Using Bi-Layer Metasurfaces and Its Application to Optical Encryption

*Hyeonhee Kim*[1,†], *Joonkyo Jung*[1,†] and *Jonghwa Shin*[1,*]

[1]Department of Materials Science and Engineering, KAIST, Daejeon 34141, Republic of Korea

[†]These authors equally contributed to this work.

[*]E-mail: qubit@kaist.ac.kr

**Abstract:** The study of optical systems with asymmetric responses has grown significantly due to their broad application potential in various fields. In particular, Janus metasurfaces are notable for their ability to control light asymmetrically at the pixel level within thin films. However, previous demonstrations were restricted to the partial control of asymmetric transmission for a limited set of input polarizations, focusing primarily on scalar functionalities. Here, we introduce optical metasurfaces consisting of bi-layer silicon nanostructures that can achieve a fully generalized form of asymmetric transmission for any possible input polarization. Their designs owe much to our theoretical model of asymmetric optical transmission in reciprocal systems that explicates the relationship between front- and back-side Jones matrices for general cases, revealing a fundamental correlation between the polarization-direction channels of opposing sides of incidence. To practically circumvent this constraint, we propose a method to partition transmission space, enabling the realization of four distinct vector functionalities within the target volume. As a proof of concept, we experimentally demonstrate polarization-direction-multiplexed Janus vector holograms generating four vector holograms. When integrated with computational vector polarizer arrays, this approach facilitates optical encryption with a high level of obscurity. We anticipate that our mathematical framework and novel material systems for generalized asymmetric transmission may pave the way for applications such as optical computations, sensing, and imaging.

## 1. Introduction

Depending on the direction of stimulus, a system of materials may respond differently. These asymmetric responses can manifest in a wide variety of systems, including electrical, optical, chemical, or mechanical devices. Such effects have captivated many researchers due to their academic intrigue and potential for many useful applications. A common yet simplistic method to achieve asymmetric responses, such as asymmetric wettability [1, 2], chemical reactivity [3, 4], or reflection [5, 6], involves constructing each side of the system from different materials or structures. This straightforward method has established a foundational basis for more intricate applications.

Beyond these basic constructs, there also exists a subtler and often surprising property of the system: the asymmetric transport of matters and energy, where the direction of incidence affects the quantity or other properties of what is being transported. A prime example of this phenomenon is the electric diode, whose unidirectional electric conduction has made them essential building blocks of modern electronic circuits and devices. Inspired by this example, numerous attempts have been made to extend this asymmetric transport phenomenon to a diverse range of matters such as liquids [2, 7, 8] or ions [9-11].

Beyond matters, the energy and information flow can also exhibit directional asymmetry, intriguing many researchers due to its unusual underlying physics and impactful potential applications. Examples include diode-like one-way transmission of elastic/acoustic waves [12-16] and heat [17-20]. In the optics community, controlling the amplitude, phase, or polarization of transmitted light depending on the direction of incidence has profound implications. Such control has been eagerly sought after for its utility in various fields of optical science and engineering. Furthermore, even greater functionality can be envisioned if the asymmetry of light transmission can be designed in a pixel-wise manner at wavelength-scale spatial resolutions over an optical interface, enabling spatially-varying light manipulation with exceptionally high resolution. The resulting thin optical systems can act as completely different devices depending on the incidence direction of light: for example, a magnifying lens with front-side illumination and a polarization camera with back-side illumination as depicted in Figure 1.

Most optical systems, except for special cases such as magneto-optic, nonlinear, or temporally modulated devices, are governed by Lorentz reciprocity [21]. A reciprocal system is often considered unsuitable for achieving asymmetric transmission because reciprocity ensures that the transmission coefficients, including their phases, remain identical when the input and output modes are swapped (i.e., reversing the direction of incidence) [22, 23]. However, the massive plurality of modes in free space presents an opportunity to achieve "apparent asymmetric transmission" without violating reciprocity. By leveraging additional degrees of freedom such as incidence angle or polarization, there are still many intriguing ways to realize asymmetric transport of energy and information within reciprocal systems. For example, consider a scenario where the incident light is restricted to a specific polarization and the system is designed such that the polarization of the transmitted light is orthogonal to that of the incident beam. Under these conditions, flipping the sample without changing the incident light results in transmission coefficients that appear completely different [24]. This is because reciprocity only dictates that the transmission for the flipped sample be the same as the original configuration when the polarization of the incident beam is also adjusted to match what was initially the output polarization. Thus, by carefully designing the optical system's response to different polarizations, asymmetric transmission can be obtained without violating Lorentz reciprocity.

Although the novel cross-polarized transmission-based approach, which utilizes orthogonal input and output polarizations for asymmetric transmission, has opened possibilities of utilizing asymmetric transmission within reciprocal systems, it has been difficult to realize using natural materials due to their limited optical anisotropy [25]. Metasurfaces, thin film-like two-dimensional arrays of nanostructures at subwavelength scale, have emerged as a transformative new optical platform. These structures offer control of light with unprecedented degrees of freedom, overcoming the limitations inherent in natural materials. Initially, metasurface-based asymmetric transmission research focused on the asymmetric transport of energy [26-29] where transmitted intensities differ for the front- and back-side illuminations. More recently, the focus has shifted towards precise phase control of transmitted electromagnetic waves. This precise phase control is very significant because it provides greater flexibility and efficiency in designing optical devices without attenuating transmitted energies. These advancements have led to the development of metasurfaces with bidirectional functionalities, also known as Janus metasurfaces, now being explored for applications such as bidirectional optical communications [30-32], holograms [30, 31, 33-42], or optical encryptions [35, 43, 44]. Despite these innovations, the question of whether optical devices with independent and arbitrary asymmetric functionalities for the front-side and back-side illuminations as described in Figure 1 can be realized has not been clearly answered, because previous demonstrations were restricted to the partial control of asymmetric transmission where limited input polarization states and scalar functionalities were of interest. Thus, achieving full control of asymmetric transmission remains a challenging task.

Here, we suggest a novel strategy for achieving full control over the asymmetric transmission of light in optical systems. We propose a generalized mathematical framework of the asymmetric transmission of an optical system, clearly revealing the relationship between transmission coefficients of the reciprocal system for the front- and back-side illuminations with respect to general elliptical polarization. This generalization enables the tailored design of systems capable of generalized asymmetric transmission for any polarization. In particular, based on our recent work that has proven that bi-layer metasurfaces can achieve complete linear control of coherent light [45], we have demonstrated thin film-like structures that fully control asymmetric transmission, allowing for the customization of both co- and cross-polarized transmission coefficients. Our research also identifies a fundamental constraint on the polarization-direction-multiplexing capability of asymmetric transmission in passive and reciprocal systems. Specifically, it is not possible to simultaneously design four arbitrary vector functionalities for four independent polarization-direction illumination conditions. To circumvent this constraint, we propose a new approach of partitioning transmission space such that independent and distinctive vector functionalities can be achieved for four different illumination conditions within the maximum target volume. As an example, we have experimentally demonstrated polarization-direction-multiplexed Janus vector holograms where unique vector holograms are generated depending on the direction of incidence and input polarization. Furthermore, we propose a novel optical encryption algorithm enabled by integrating our Janus metasurfaces with computational vector polarizer arrays. This integration significantly enhacnes the security level of metasurface-based optical encryptions, presenting new possibilities in secure communication technologies.

## 2. Results

### 2.1. Generalization of asymmetric transmission

For the asymmetric transmission in a reciprocal system, we focus on orthogonal input–output polarization as an additional degree of freedom. The effect of an optical system on the amplitude, phase, and polarization of the incident light can be effectively captured using the Jones matrix [24, 45-48], which relates the input and output lights through matrix multiplication with respect to two polarization states as bases. According to reciprocity, the

Jones matrices for the front- and back-side illuminations are correlated [24]. For the *x*- and *y*-polarization bases, if the front-side Jones matrix of an optical system is expressed as $T_{xy}^{\mathrm{f}} = [A, B; C, D]$, then the back-side Jones matrix when the system is flipped becomes $T_{xy}^{\mathrm{b}} = [A, -C; -B, D]$ where the subscript *xy* represents the *x*- and *y*-polarization bases and the superscripts f and b represent the front- and back-side illuminations, respectively. Similarly, for the right- and left-handed circular polarization (RCP and LCP) bases, if the front-side Jones matrix of a system is given as $T_{RL}^{\mathrm{f}} = [A', B'; C', D']$, then the back-side Jones matrix becomes $T_{RL}^{\mathrm{b}} = [A', C'; B', D']$ where the subscript *RL* represents the RCP and LCP bases and the superscripts f and b represent the front- and back-side illuminations, respectively. Note that in both cases, while diagonal elements (co-polarized transmission coefficients) remain the same, off-diagonal elements (cross-polarized transmission coefficients) are interchanged, with or without a sign conversion.

Most previous studies have utilized the interchange of cross-polarized transmission coefficients to achieve asymmetric transmission with specific cross-polarized input–output channels, such as *x*-polarized input and *y*-polarized output [31-38, 49], or RCP input and LCP output [30, 39-42]. By strategically designing these off-diagonal elements and minimizing the influence of diagonal elements, researchers have developed devices that offer different functionalities for the front- and back-side illuminations in these specific cross-polarization channels. However, the relationship between the front- and back-side Jones matrices under arbitrary elliptical polarization bases is not as straightforward as merely interchanging off-diagonal elements [24]. This complexity has resulted in limited exploration of asymmetric transmission relative to general elliptical polarization bases, and a general expression for this phenomenon remains undefined. This gap has hindered efforts to define the clear boundaries of what kinds of asymmetric functionalities can or cannot be integrated within a single device.

Moreover, while prior studies have mainly focused on manipulating only off-diagonal elements of the Jones matrix, a deliberate and harmonious design of both diagonal and off-diagonal elements is required to take full advantage of the vector nature of electromagnetic fields. Although there is one clever exception where the co- and cross-polarized transmissions are concurrently utilized [43], this example still falls short of achieving full control over asymmetric transmission. Therefore, it is expected that a generalized understanding of asymmetric transmission and the simultaneous designability of co- and cross-polarized transmissions can create an intriguing avenue for further inquiry and innovation in the field of metasurface-enabled asymmetric transmission.

To address these issues, we have established a generalized form for representing the Jones matrices for the front-and back-side illuminations with elliptical polarization bases, consistent with the case of the *x*- and *y*-polarization bases. This involves simply interchanging off-diagonal elements with a sign conversion. It is important to note the distinction between reversing the direction of illumination and flipping the optical system itself; the latter scenario is akin to reversing the direction of illumination accompanied by a coordinate rotation (See Figure S1, Supporting Information for the details).

For our generalization, we consider an arbitrary orthogonal pair of elliptical polarizations, $\lambda$- and $\lambda_\perp$-polarizations, defined as $|\lambda\rangle = [\cos\chi; \sin\chi\, \mathrm{e}^{i\theta}]$ and $|\lambda_\perp\rangle = [\sin\chi; -\cos\chi\, \mathrm{e}^{i\theta}]$, respectively. With these polarizations as the bases, the front-side Jones matrix of an optical system is expressed as $T_{\lambda\lambda_\perp}^{\text{f}} = [a, b; c, d]$ where the subscript $\lambda\lambda_\perp$ represents the $\lambda$- and $\lambda_\perp$-polarization bases. Figure 2(a) schematically illustrates this matrix as a linear network connecting the input and output ports (from left to right) relative to these bases, with matrix elements corresponding to transmission coefficients for co- and cross-polarization channels.

By changing the bases to the *x*- and *y*-polarizations, the front-side Jones matrix can be rearranged as

$$T_{xy}^{\text{f}} = \begin{bmatrix} A & B \\ C & D \end{bmatrix} = [|\lambda\rangle \quad |\lambda_\perp\rangle] \begin{bmatrix} a & b \\ c & d \end{bmatrix} \begin{bmatrix} \langle\lambda| \\ \langle\lambda_\perp| \end{bmatrix}. \tag{1}$$

At this point, reciprocity tells us that if the direction of incidence is reversed and the polarization states under consideration are the time-reversed or the conjugate forms of those used for front-side illumination case, the system exhibits the same transmission behaviors. Thus, the Jones matrix for this reciprocal scenario becomes $T_{\lambda^*\lambda_\perp^*}^{\text{r}} = [a, c; b, d]$, essentially the transpose of the front-side Jones matrix relative to the $\lambda$- and $\lambda_\perp$-polarization bases, where the superscript r represents the reciprocal scenario and the subscript $\lambda^*\lambda_\perp^*$ represents the $\lambda^*$- and $\lambda_\perp^*$-polarization bases, defined as $|\lambda^*\rangle = [\cos\chi; \sin\chi\, e^{-i\theta}]$ and $|\lambda_\perp^*\rangle = [\sin\chi; -\cos\chi\, e^{-i\theta}]$. Similarly, this Jones matrix can be rearranged relative to the *x*- and *y*-polarization bases as

$$T_{xy}^{\text{r}} = \begin{bmatrix} A & C \\ B & D \end{bmatrix} = [|\lambda^*\rangle \quad |\lambda_\perp^*\rangle] \begin{bmatrix} a & c \\ b & d \end{bmatrix} \begin{bmatrix} \langle\lambda^*| \\ \langle\lambda_\perp^*| \end{bmatrix}. \tag{2}$$

As indicated by Equation (1) and (2), the Jones matrices for the front-side illumination and its reciprocal scenario are simply transposes of each other only when represented in conjugate polarization bases. Previous studies have failed to show such a straightforward relationship for general elliptical polarization bases, as they have used the same polarization bases for the front- and back-side illuminations.

However, in this fixed coordinate, the definition of handedness of polarization states depends on the direction of incidence due to the reversed propagation direction. For example, RCP is represented as $\left[\frac{1}{\sqrt{2}}; \frac{i}{\sqrt{2}}\right]$ for the front-side illumination and $\left[\frac{1}{\sqrt{2}}; \frac{-i}{\sqrt{2}}\right]$ for the reversed back-side illumination. To avoid this confusion, we rotated the coordinate around *x*-axis for the back-side illumination, ensuring consistent notation of polarization states regardless of the direction of incidence. Since this is effectively the same situation as flipping the system itself, throughout this study, we equate flipping of the system with reversing the direction of illumination accompanied by the rotation of coordinate around *x*-axis for the back-side illumination (See Figure S1, Supporting Information for the comparisons). Accordingly, the back-side Jones matrix is obtained by multiplying the matrix $Q = [1,0; 0, -1]$ on both sides of the transpose of the Jones matrix of the reciprocal scenario, which is given as

$$T_{xy}^{\text{b}} = QT_{xy}^{\text{r}}Q = [|\lambda'\rangle \quad |\lambda_\perp'\rangle] \begin{bmatrix} a & -c \\ -b & d \end{bmatrix} \begin{bmatrix} \langle\lambda'| \\ \langle\lambda_\perp'| \end{bmatrix} \tag{3}$$

where $|\lambda'\rangle = [\cos\chi; -\sin\chi\, e^{-i\theta}]$ and $|\lambda_\perp'\rangle = [-\sin\chi; -\cos\chi\, e^{-i\theta}]$ are the newly introduced polarizations (see Text S1, Supporting Information for the details). Note that by changing the bases of Equation (3), the back-side Jones matrix relative to the $\lambda'$- and $\lambda_\perp'$-polarization bases is given as $T_{\lambda'\lambda_\perp'}^{\text{b}} = [a, -c; -b, d]$, obtained by simply interchanging off-diagonal elements of the front-side Jones matrix relative to the $\lambda$- and $\lambda_\perp$-polarization bases with a sign conversion as schematically shown in Figure 1(a). Thus, we have generalized the relation of polarization bases for the asymmetric transmission such that one can always find

two orthogonal polarization bases for the front- and back-side illuminations, where their co-polarized transmission coefficients remain the same, and cross-polarized transmission coefficients are interchanged with a sign conversion. This generalization clearly shows the performance boundary of asymmetric transmission of passive and reciprocal optical systems and provides an intuitive mathematical framework to design various asymmetric optical devices in later sections.

In Figure 2(b), the polarizations of $|\lambda\rangle$, $|\lambda_\perp\rangle$, $|\lambda'\rangle$ and $|\lambda'_\perp\rangle$ are represented on the Poincaré sphere with their polarization ellipses and the Stokes parameters. Note that two points on the Poincaré sphere for $|\lambda\rangle$ and $|\lambda'\rangle$ (likewise, $|\lambda_\perp\rangle$ and $|\lambda'_\perp\rangle$) are mirror symmetric to each other with respect to the $S_1$-$o$-$S_3$ plane. This symmetry means their polarization ellipses have the same ellipticity and handedness, but the major axes are rotated in the opposite direction. This geometric relationship provides a key insight: for any orthogonal pair of polarization bases on the $S_1$-$o$-$S_3$ plane, the front- and back-side Jones matrices follow the simple relation of interchanging off-diagonal terms with a sign conversion for the same polarization bases regardless of the direction of incidence. The representative examples are the $x$- and $y$-polarization bases or RCP and LCP polarization bases (in this case, $|\lambda'_\perp\rangle$ is defined as LCP with a $\pi$ phase delay). One thing to note is that the orientation of this mirror plane can be modified by rotating the axis along which the system is flipped (see Text S1 and Figure S2, Supporting Information for the details). Therefore, by properly choosing the axis of flipping, any orthogonal pair of polarizations bases can have the simply related front- and back-side Jones matrices relative to the same polarization bases.

Recent advancements of dielectric metasurfaces have enabled the realization of arbitrary Jones matrices using bi-layer arrays with supercell structures, called universal metasurfaces, as in Figures 2(c) and 2(d) [45]. To demonstrate the proposed generalized asymmetric transmission, we randomly selected an orthogonal pair of polarization bases ($\chi = 2\pi/5$ and $\theta = \pi/6$) and designed universal metasurfaces working at a wavelength of 915 nm with a randomly chosen Jones matrix (see Experimental section and Text S2, Supporting Information for the details). Figure 2(e) shows the Jones matrices retrieved from simulations for the front- and back-side illuminations. As expected, the retrieved diagonal elements are identical and the retrieved off-diagonal elements are interchanged with a sign conversion.

Furthermore, we experimentally demonstrated the generalized asymmetric transmission by designing four different holograms for four independent co- and cross-polarization transmission channels, defined with $\chi = \pi/6$ and $\theta = \pi/4$. The required phase profiles were designed based on gradient-descent optimization method [50, 51] (see Experimental section for the details). As in Figure 2(f), for the front-side illumination with $\lambda$- and $\lambda_\perp$-polarizations, four different images—a dog, a cat, a sheep and a rat—were successfully measured for each polarization channel (see Experimental section for the details). On the contrary, for the back-side illumination with $\lambda'$- and $\lambda'_\perp$-polarizations, while the same images—the dog and the rat—were measured through the co-polarization channels, the measured images for the cross-polarization channels—the cat and the sheep—were interchanged. Additionally, when the metasurface is illuminated from the front-side with $\lambda'$- and $\lambda'_\perp$-polarizations, the overlapped images are generated since $\lambda'$- and $\lambda'_\perp$-polarizations can be regarded as the mixture of $\lambda$- and $\lambda_\perp$-polarizations (See Figure S4, Supporting Information). Note that all images for the back-side illumination were flipped upside down because the phase profiles of holograms were also flipped upside down when flipping the metasurface.

## 2.2. Polarization-direction-multiplexed Janus vector holograms

The generalized form derived in the previous section leads to a fundamental constraint on asymmetric transmission for passive and reciprocal systems. Let us consider a general optical system characterized by the Jones matrix given as $T^{\text{f}}_{\lambda\lambda_\perp} = [a, b; c, d]$. When this system is illuminated from the front-side with $\lambda$- and $\lambda_\perp$-polarization, the output field $\boldsymbol{E}$ is given as $\boldsymbol{E}^{\lambda}_{\lambda\lambda_\perp} = [a; c]$ and $\boldsymbol{E}^{\lambda_\perp}_{\lambda\lambda_\perp} = [b; d]$ where the superscript $\lambda(\lambda_\perp)$ represents the input $\lambda(\lambda_\perp)$-polarization and the subscript represents the bases, respectively. Conversely, when the system is illuminated from the back-side with $\lambda'$- and $\lambda'_\perp$-polarization, the output field $\boldsymbol{E}$ is given as $\boldsymbol{E}^{\lambda'}_{\lambda'\lambda'_\perp} = [a; -b]$ and $\boldsymbol{E}^{\lambda'_\perp}_{\lambda'\lambda'_\perp} = [-c; d]$ where the superscript $\lambda'(\lambda'_\perp)$ represents the input $\lambda'(\lambda'_\perp)$-polarization and the subscript represents the bases, respectively.

These four output fields share common elements with each other, demonstrating their interdependence. This interdependence highlights a lack of degrees of freedom necessary to achieve four independent vector field distributions, each requiring independent control over both components of the output electric fields under four different polarization-direction conditions of illumination. In addition, this analysis reveals that the co-polarized transmission coefficients remain unchanged when the direction of illumination is reversed; in other words, they are symmetric. Due to this symmetry, the manipulation of polarization has been of less interest in the field of Janus metasurfaces. Consequently, most prior studies have focused solely on scalar functionalities through the cross-polarized transmission channel where the polarization profile of output fields is homogeneous and fixed to be orthogonal to input polarization.

However, controlling the polarization of electromagnetic waves is fundamentally significant for enabling a variety of polarization-related applications such as optical computations [52-54], quantum optics [55-59] or optical imaging [60-63]. To address the constraint associated with Janus metasurfaces, we suggest a novel polarization-direction multiplexing approach. This technique allows a single metasurface to operate independently under different polarization-direction conditions of illumination. The suggested polarization-direction-multiplexed Janus vector holograms can implement four independent asymmetric vector functionalities where the polarization profiles of output fields can be designed to vary spatially. These Janus vector holograms are based on the spatial partitioning of transmission space into upper and lower half-spaces where only the upper half-space is used as the active target spatial channel, while the lower half-space is blocked or discarded as illustrated in Figure 3(a).

This approach maximizes the available target spatial volume for vector functionalities up to half of transmission space by utilizing the aforementioned spatial flipping of output fields accompanied by flipping of the system. For example, if the element $a(x, y)$ of the Jones matrix is designed to generate an image in the upper half-space for the front-side illumination, then for the back-side illumination, where the system is flipped, this image is shifted to the lower half-space, leaving the upper half-space empty. On the contrary, the same element

$a(x, y)$ can be designed to generate an image in the upper half-space for back-side illumination, then for the front-side illumination, this image is shifted to the lower half-space. Thus, by superposing two such functions in the upper-half space for both directions of illumination and discarding information in the lower-half space, $a(x, y)$ can be designed to achieve different functions in the target spatial channel (i.e., upper half-space) for the front- and back-side illuminations without crosstalk. Similarly, by designing other elements—$b(x, y)$, $c(x, y)$, and $d(x, y)$—it is possible to realize four independent vector functionalities within a single optical system. Furthermore, if one can design arbitrary vector functionalities for the input $\lambda'$- and $\lambda'_\perp$-polarizations, it implies that arbitrary vector functionalities can be designed for any other orthogonal pair of input polarizations; in other words, input polarizations for the front- and back-side illuminations do not have to be related, and any input polarizations for the back-side illumination can be considered, represented as $\eta$- and $\eta_\perp$-polarizations in Figure 3(a).

To experimentally demonstrate our proposed polarization-direction-multiplexed Janus vector holograms, we meticulously designed and fabricated Janus metasurfaces capable of generating four different vector holographic images. Specifically, the metasurface produces images of a butterfly and a grasshopper under the front-side illumination with $\lambda$- and $\lambda_\perp$-polarizations, respectively. On the other hand, the images of a ladybug and a beetle are generated for the back-side illumination with $\eta$- and $\eta_\perp$-polarizations, as described in Figure 3(a). To maximize efficiency, each constituent unit cell was designed to have a unitary Jones matrix, and then the Janus metasurface was optimized using gradient-descent optimization (see Experimental section for the details of design and optical characterization).

The calculated efficiencies of the four vector holographic images are around 31.9%, 52.8%, 44.0% and 34.4%, respectively, with an ideal metasurface of unity transmission assumed. Throughout this work, we visualize polarization and intensity using false color by mapping the normalized Poincaré sphere into the CIELAB color space, as shown in Figure 3(b). In this visualization, the azimuth and elevation angles represent the polarization and the radius represents the intensity such that the maximum intensity is represented as a radius of one.

Figure 3(c) presents both the target and the measured holographic images in the upper half-space (see Figure S5, Supporting Information for the optimized and measured images in the full transmission space). As can be seen, the target and measured images show good agreement. Degradation and speckle noise in the measured images are mainly attributable to imperfections of fabrication and measurement, as well as characteristics of the laser source used.

## 2.3. Integration of Janus metasurfaces with state-of-the-art hologram techniques

Modern hologram techniques have continuously evolved, enhancing the ability to store more information within a single device. Our proposed Janus vector holograms can integrate seamlessly with such further advanced metasurface-based hologram techniques. As representative examples, we have adopted two state-of-the-art hologram techniques: multi-plane holograms [45, 64-67] and multichannel holograms with nonseparable polarization transformation [50], as schematically described in Figure 4(a) and 4(b).

Unlike traditional far-field holograms that generate a single target image at the far-field plane,

multi-plane holograms produce multiple images at various positions within the Fresnel diffraction regime. This ability to continuously modify the diffracted pattern based on the propagation of waves makes multi-plane holography a prominent approach for creating three-dimensional holographic images [45, 64, 67]. Our proposed Janus metasurfaces can further enhance the ability to store information in multi-plane holograms by adding additional degrees of freedom through polarization-direction of illumination. We experimentally demonstrated Janus metasurfaces that can store multiple vector holographic images in both the far-field and the Fresnel diffraction regimes for each illumination condition (see Experimental section for the details). Figure 4(a) illustrates the schematic of the designed device for a specific incidence condition (the front-side illumination with $\lambda$-polarization). Specifically, the device is designed to generate vector holographic images of a butterfly, a grasshopper, a ladybug, and a beetle in the far-field regime, and a dog, a sheep, a cat, and a rat in the Fresnel regime for four different incidence conditions: front-side illumination with $x$- and $y$-polarizations, and back-side illumination with diagonal and anti-diagonal polarizations, respectively.

In Figure 4(c), the measured vector images are presented for four different incidence conditions. The measured results are all in excellent agreement with the target (see Figure S6, Supporting Information). Although the same input polarization was considered for the holographic images in the far-field and Fresnel diffraction regimes, the system can be designed to operate with distinct input polarizations for different imaging planes. Note that the number of stored images can be increased at the cost of the quality and complexity of the holographic images.

Conventionally, polarization-multiplexed holograms are designed to generate just two target images for an orthogonal pair of input polarizations. In such configurations, when the input polarization differs from the designed ones, the overlapped image of two targets is usually observed. However, leveraging the vector nature of electromagnetic fields can significantly expand this capability, enabling the creation of multiple target images for various input polarizations beyond an orthogonal pair of polarizations within a single device. This advanced approach, known as multichannel holograms with nonseparable polarization transformation [50], allows for more complex image storage.

Similarly, our Janus metasurfaces can further enhance the capability to store information in multichannel holograms. We experimentally demonstrated direction-multiplexed multichannel holograms with nonseparable polarization transformation using Janus metasurfaces (see Experimental section for the details). Figure 4(b) presents the schematic of the designed device under front-side illumination. Specifically, the device was designed to generate the letters "A", "B", "C", and "D" for the front-side illumination with $x$-, $y$-, diagonal and anti-diagonal polarizations, and the Greek letters "α", "β", "γ", and "δ" for the back-side illumination with $x$-, $y$-, RCP, and LCP polarizations, respectively. In Figure 4(d), the measured intensity profiles show good agreement with the optimized images (see Figure S7, Supporting Information). Likewise, note that the number of stored images can be increased at the cost of the quality and complexity of the holographic images.

## 2.4. Optical encryption based on Janus vector holograms with high-security level

Numerous studies have explored the application of metasurfaces in optical encryption, recognizing their potential to significantly enhance security levels. Metasurfaces are particularly advantageous for optical encryption due to two main aspects. First, their capacity for multichannel operation, utilizing polarization [68-72], wavelength [72-74], orbital angular momentum [73, 75, 76], or direction [35, 43, 44] as multiplexing methods, allows for the selective access to ciphertexts. This selective access ensures that users can only view ciphertexts when specific, predetermined conditions are satisfied, thereby boosting security in an efficient and effective manner. Second, the high degrees of freedom that metasurfaces offer in controlling polarization, phase, and amplitude enable the creation of highly complex ciphertexts, further enhancing security levels.

Polarization is especially valuable in this context because it is inherently multi-dimensional, allowing for the storage of a huge amount of information or the concealment of data awaiting decryption. By making the most of our Janus metasurfaces, we introduce an advanced optical encryption scheme with a very high-security level. In this scheme, ciphertexts are accessible only under the correct conditions of input polarization and direction of illumination, and they require a complex polarization filtering process for decryption.

As schematically described in Figure 5(a), combined with computational vector polarizers (VPs) comprising arrays of pixelated polarizers, our proposed polarization-direction-multiplexed Janus holograms enable optical encryption with a high-security level. The encryption mechanism can be geometrically understood based on the Poincaré sphere, as shown in Figure 5(b). A polarizer typically allows the passage of its desired polarization while blocking its undesired, orthogonal counterpart; for example, an *x*-polarizer transmits *x*-polarization and blocks *y*-polarization. When an arbitrary polarization is introduced, only the energy corresponding to the desired polarization can pass through, which is also known as Malus' law [77].

On the Poincaré sphere, an orthogonal pair of desired and undesired polarizations of polarizers makes up an axis, e.g., $S_1$-axis for the *x*-polarizer. Then, we can consider a great circle on the sphere perpendicular to the axis given by the polarizer, e.g., the great circle on $S_2$-*o*-$S_3$-plane for the *x*-polarizer. This great circle divides the Poincaré sphere into two hemispheres: one closer to the desired polarization and the other closer to the undesired polarization. While polarizations on the great circle have the same amount of energy of the desired and undesired polarizations, the hemisphere closer to the desired polarization contains a larger amount of the desired polarization compared to the undesired one, and the other hemisphere contains a larger amount of the undesired polarization. Thus, polarizations on these two hemispheres can be considered to represent polarizations that show "relatively high" and "relatively low" intensities after passing through the polarizer, respectively, which are represented as "1" and "0" states in our work, making this division useful for optical encoding.

Similarly, we can expand this concept. If we consider three polarizers ($p_1$, $p_2$ and $p_3$) whose axes on the Poincaré sphere are mutually perpendicular (e.g., *x*-, diagonal and RCP polarizers), the Poincaré sphere is divided into eight octants as described in Figure 5(b). Each octant has its own characteristic intensity pattern with respect to the given three polarizers. For example, as described in the inset of Figure 5(b), polarizations in the "011" octant show the intensity pattern of "relatively low" for the first *x*-polarizer, "relatively high" for the second diagonal polarizer, and "relatively high" for the third RCP polarizer. Therefore, by using polarization of electromagnetic waves, we can store 3-bit information in this configuration. To ensure uniform intensity differences between the "1" and "0" states for the three polarizers, we used only the eight polarizations selected as the vertices of the inscribed cube, which are aligned with the three axes of the polarizers. Note that by considering more polarizers, information with higher density could be stored in polarization of light using a more finely segmented Poincaré sphere.

By using polarization as an information carrier, three pixelated binary images can be encrypted into a single vector hologram with uniform intensity and complex polarization profiles in a pixelated pattern, as in Figure 5(a). For every pixel, three polarizers—$p_1$, $p_2$, and $p_3$—are randomly selected; then, three computational VPs are defined as arrays of these three polarizers for all pixels. When this vector hologram is computationally decoded by using the appropriate VPs as the keys, three images can be retrieved as represented in Figure 5(a), highlighting the hologram's encryption capabilities.

As mentioned, the Janus metasurfaces' capability of generating different vector holograms for different polarization-direction illumination conditions enables optical encryption with a very high-security level when combined with our novel encryption mechanism. Successful decryption occurs only when all the incidence direction, the input polarization, and the appropriate VP are satisfied, allowing the correct information to be extracted from a metasurface. To experimentally demonstrate this capability, we designed and fabricated Janus vector holograms that can produce four different vector holograms generating 8-by-8 and 10-by-10 pixelated polarization patterns for the front- and back-side illuminations with $x$- and $y$-polarizations. These polarization patterns, or ciphertexts, can be decoded by predefined VPs as decoding keys.

The fabricated samples were measured, and the hidden images were successfully retrieved from measurements using the predetermined computational VPs (see Experimental section for the details). Note that all the hidden images were independently designed without any crosstalk. Figure 5(c) presents the measured and retrieved results of the holograms generating 10-by-10 pixelated polarization patterns (see Figure S8, Supporting Information for the results of the holograms generating 8-by-8 pixelated patterns and Figure S9, Supporting Information for the raw measured data and optimized polarization patterns).

In the left panels of Figure 5(c), the raw measured data, including the Stokes parameters and polarization patterns for the front- and back-side illuminations with the designed input polarization of $x$- and $y$-polarizations, are displayed. These raw measurements cleverly conceal any discernible clues about the hidden images, demonstrating the robustness of our encryption method. On the contrary, when the correct computational VPs were applied, all the hidden images were well extracted as shown in the right panels, and all these results are in good agreement with the ground truth binary images (see Figure S10, Supporting Information). The hidden images can be utilized in many ways. For example, each hidden image can play a role of true information by itself; some images represent true information, while others serve as decoy images; each hidden image stores a fragment of the complete information.

Our proposed optical encryption algorithm can be securely implemented using a dual-channel method for secret transmission. For example, through the first channel, a metasurface is physically delivered, while through the second channel, detailed instructions on the necessary incidence conditions are delivered. Correct decryption of the information occurs only when the metasurface, the instructions, and the decryption key (the VP) are precisely matched. Even if one or two of those components are leaked by eavesdroppers, the encrypted information remains secure. Note that the level of security can be more enhanced when more pixels or more modulations of polarization are further considered [68, 77].

## 3. Conclusion

We have introduced a new perspective on asymmetric transmission using a generalized formulation where the roles of co- and cross-polarized transmission coefficients for the front- and back-side illuminations are clearly correlated based on two related polarization bases. We numerically and experimentally demonstrated that thin film-like structures can achieve complete control over asymmetric transmission by designing metasurfaces with tailored co- and cross-polarized transmission coefficients.

We then experimentally showed that the simultaneous designability of co- and cross-polarized transmission enabled polarization-direction-multiplexed Janus vector holograms, inscribing four different vector functionalities within a single metasurface. Furthermore, we seamlessly integrated our proposed Janus vector holograms with modern advanced hologram techniques, such as multi-plane holograms and multichannel holograms with nonseparable polarization transformation. Lastly, we introduced a novel optical encryption scheme where our Janus metasurfaces, combined with the computational VPs, provide a highly enhanced level of security.

Looking to the future, we see considerable potential for further enhancing the performance of Janus metasurfaces by integrating them with various active metasurface techniques based on electrical modulation [78, 79], mechanical reconfiguration [80-82], or phase-change materials [35, 44, 83]. These active metasurfaces inherently allow for asymmetric transmission when external stimuli are applied. Interestingly, the asymmetric transmission facilitated by active metasurfaces can be qualitatively different from that by Janus metasurfaces; it can rely on the dynamic optical properties of the system, whereas the latter depend on static optical properties. Therefore, we anticipate that Janus metasurfaces and active metasurfaces can be combined in a complementary manner.

Furthermore, recent advancements in nonlinear metasurfaces [84, 85] and space-time metasurfaces [86, 87] are noteworthy because they could provide non-reciprocal asymmetric transmission. These advancements could potentially enable the full utilization of transmission space for polarization-direction-multiplexed Janus vector holograms when combined with our Janus metasurfaces. Overall, we anticipate that the extended functionalities offered by our Janus metasurfaces may find applications in many other scientific and engineering fields.

## 4. Experimental Section

### *Design of Universal Metasurfaces in Figure 2*

The unit cell structure of universal metasurfaces, operating at a wavelength of 915 nm, was composed of an SiO$_2$ substrate, Si nanoposts, and a SU-8 spacer layer, as shown in Figure 2(c). The lateral size was set at 450 nm, and the heights of Si nanoposts in the lower and upper layers and the SU-8 spacer layer were 790 nm, 700 nm and 1300 nm, respectively. For fabrication feasibility, the lengths of the major and minor axes of the nanoposts were restricted to a range of 100 nm to 350 nm. The complex refractive indices of the constituent materials were assumed to be 1.45 for SiO$_2$, 3.61 + 0.0066i for Si, and 1.56 for SU-8 polymer. For numerical demonstration, commercial FDTD software from Ansys Lumerical Inc. was utilized. We randomly generated an arbitrary matrix and decomposed it to obtain related structural parameters of the universal metasurfaces (see Text S2, Supporting Information for the details). Using the obtained structural parameters, the universal metasurfaces were designed with subcells of 11-by-11 clusters of bi-layer structures.

*Fabrication of Bi-Layer Metasurfaces*

The first layer of amorphous Si, with a thickness of 790 nm, was deposited on quartz glass using plasma-enhanced chemical vapor deposition (PECVD). For electron-beam (e-beam) lithography, an adhesion layer (AR 300-80, Allresist), e-beam resist (AR-P 6200.04, Allresist), and conductive polymer (AR-PC 5090.02, Allresist) were spin-coated. Then, alignment-marks were patterned using e-beam lithography, followed by metal deposition of Cr with 20 nm thickness and Au with 50 nm thickness using e-beam evaporation and lift-off process. For the patterning of nanopost structures on the lower layer, the e-beam lithography process was employed. An alumina hard mask with a thickness of 60 nm was deposited by e-beam evaporation, and the lift-off process was performed. To achieve vertical sidewalls and a high aspect ratio, the deep reactive ion etching (RIE) technique, known as pseudo-Bosch dry etching, was utilized. Then, the first layer was coated by spin-coating SU-8 polymer with a 1.7 µm thickness to encapsulate the first layer and provide a flattened surface for the deposition of the second layer. The second layer of amorphous Si, with a thickness of 650 nm, was deposited by radio-frequency sputtering. Nanopost structures on the second layer were patterned using a process similar with that of the first layer. For experimental demonstration, pixels of Janus metasurfaces were composed of 3-by-3 clusters of bi-layer structures.

*Design of Scalar Holograms in Figure 2*

All holograms demonstrated in this work were optimized using gradient-descent optimization (See Text S3, Supporting Information for the details). For optimization of a multifunctional device with various incidence conditions, the total loss function was set as $L^{\text{tot}} = \sum_i L^{(i)}$, where $L^{\text{tot}}$ is the total loss function, and $L^{(i)}$ is the loss function of the $i$-th function of a metasurface. For the optimization of scalar holograms, we used the loss function given as $L^{(i)} = \frac{1}{N^2} \sum_p \sum_q \left( \left| E_{pq,x}^{(i)} t_x^{*,(i)} + E_{pq,y}^{(i)} t_y^{*,(i)} \right|^2 - I^{(i)} \right)^2$, where $N$ is the number of spatial points in the $k_x$ and $k_y$ domain in the far-field plane, the subscripts $p$ and $q$ represent the indices of spatial points in the $k_x$ and $k_y$ domain, $E_x^{(i)}$ and $E_y^{(i)}$ are the $x$- and $y$-components of the output electric field in the far-field regime for the $i$-th incidence condition, $I^{(i)}$ is the intensity profile of the $i$-th holographic image, the superscript * represents conjugation, and $t_x^{(i)}$ and $t_y^{(i)}$ represent the $x$- and $y$-components of the target polarization of the $i$-th holographic image. Intuitively, the first term of the loss function can be understood as a term for inserting a polarizer to pass the desired polarization. Then, through the optimization, this intensity of the desired polarization gets closer to the target intensity profile.

*Design of Polarization-Direction-Multiplexed Janus Vector Holograms in Figure 3 and Figure 5*

For optimization of Janus vector holograms that generate four vector holographic images in the far-field regime with spatially varying polarization for different incidence conditions, we used the loss function given as $L^{(i)} = \frac{1}{N^4}\sum_p \sum_q \left[\left(\left|E^{(i)}_{pq,x}\right|^2 - \left|t^{(i)}_{pq,x}\right|^2\right)^2 + \left(\left|E^{(i)}_{pq,y}\right|^2 - \left|t^{(i)}_{pq,y}\right|^2\right)^2 + \left|E^{(i)}_{pq,y}t^{(i)}_{pq,x} - E^{(i)}_{pq,x}t^{(i)}_{pq,y}\right|^2\right]$, where $N$ is the number of spatial points in the $k_x$ and $k_y$ domain in the far-field plane, the subscripts $p$ and $q$ represent the indices of spatial points in the $k_x$ and $k_y$ domain, $E^{(i)}_x$ and $E^{(i)}_y$ are the x- and y-components of the output electric field in the far-field regime for the *i*-th incidence condition, and $t^{(i)}_x$ and $t^{(i)}_y$ represent the x- and y-components of the target output field of the *i*-th holographic image. The first and second terms of the loss function can be intuitively understood as terms for matching the amplitude of x- and y-polarization components, and the third term as a term for matching the relative phase between them. The total loss function was given as $L^{\text{tot}} = \sum_i L^{(i)}$. It is important to note that since only the upper half-space was of interest, only the gradient information of the upper half-space was considered during the optimization to maximize efficiency through the target spatial channel.

*Design of Multi-Plane Janus Vector Holograms in Figure 4*

For optimization of multi-plane Janus vector holograms that generate eight vector holographic images in the far-field and the Fresnel diffraction regimes, we used two kinds of loss functions. The first kind, used to optimize holograms for the far-field regime, was given as $L^{(i)}_1 = \frac{1}{N^4}\sum_p \sum_q \left[\left(\left|E^{(i)}_{pq,x}\right|^2 - \left|t^{(i)}_{pq,x}\right|^2\right)^2 + \left(\left|E^{(i)}_{pq,y}\right|^2 - \left|t^{(i)}_{pq,y}\right|^2\right)^2 + \left|E^{(i)}_{pq,y}t^{(i)}_{pq,x} - E^{(i)}_{pq,x}t^{(i)}_{pq,y}\right|^2\right]$, where $N$ is the number of spatial points in the $k_x$ and $k_y$ domain in the far-field plane, the subscripts $p$ and $q$ represent the indices of spatial points in the $k_x$ and $k_y$ domain, $E^{(i)}_x$ and $E^{(i)}_y$ are the x- and y-components of the electric field in the far-field regime for the *i*-th incidence condition, and $t^{(i)}_x$ and $t^{(i)}_y$ represent the x- and y-components of the target output field of the *i*-th holographic image in the far-field regime. The second kind of loss function, used to optimize the holograms for the Fresnel diffraction regime, was given as $L^{(j)}_2 = \sum_a \sum_b \left[\left(\left|E'^{(j)}_{ab,x}\right|^2 - \left|t'^{(j)}_{ab,x}\right|^2\right)^2 + \left(\left|E'^{(j)}_{ab,y}\right|^2 - \left|t'^{(j)}_{ab,y}\right|^2\right)^2 + \left|E'^{(j)}_{ab,y}t'^{(j)}_{ab,x} - E'^{(j)}_{ab,x}t'^{(j)}_{ab,y}\right|^2\right]$, where the subscripts $a$ and $b$ represent the indices of spatial points in the real domain of the Fresnel diffraction regime, $E'^{(j)}_x$ and $E'^{(j)}_y$ are the x- and y-components of the electric field in the Fresnel diffraction regime for the *j*-th incidence condition and $t'^{(j)}_x$ and $t'^{(j)}_y$ represent the x- and y-components of the target output field of the *j*-th holographic image in the Fresnel diffraction regime. The total loss function was given as $L^{\text{tot}} = \sum_i L^{(i)}_1 + \sum_j L^{(j)}_2$. Note that only the gradient information of the upper half-space was considered during the optimization to maximize efficiency through the target spatial channel.

## *Design of Multichannel Holograms with Nonseparable Polarization Transformation in Figure 4*

For optimization of multichannel holograms with nonseparable polarization transformation, we used the loss function given as $L^{(i)} = \frac{1}{N^4} \sum_p \sum_q \left( \left|E^{(i)}_{pq,x}\right|^2 + \left|E^{(i)}_{pq,y}\right|^2 - I^{(i)} \right)^2$, where $N$ is the number of spatial points in the $k_x$ and $k_y$ domain in the far-field plane, the subscripts $p$ and $q$ represent the indices of spatial points in the $k_x$ and $k_y$ domain, $E^{(i)}_x$ and $E^{(i)}_y$ are the $x$- and $y$-components of the output electric field in the far-field regime for the $i$-th incidence condition and $I^{(i)}$ is the intensity profile of the $i$-th holographic image. The total loss function was given as $L^{\text{tot}} = \sum_i L^{(i)}$.

## *Optical Characterization*

Scalar holograms and multichannel holograms with nonseparabale polarization transformation were characterized using the setup shown in Figure S12, Supporting Information. The polarization of incident light was controlled using a half-wave plate and a quarter-wave plate. After passing through the device, the Fourier plane was formed at the back-focal plane (BFL) of an objective lens (MPLFLN 10x, Olympus). To facilitate measurement, the Fourier plane was projected onto a charge-coupled device (CCD) camera (CS505MU, Thorlabs) using a 4-*f* system (LSB04, Thorlabs) with a focal length of 200 mm. For measuring scalar holograms, a linear polarizer and a quarter-wave plate were placed in front of the CCD camera to pass only the desired elliptical polarization. Vector holograms were characterized by Stokes parameters, which can be determined by four intensity measurements with the predefined configurations of a linear polarizer and a quarter-wave plate in front of the CCD camera [45, 88]. Holographic images in the Fresnel regime can be measured directly by measuring the Fresnel diffraction regime using a tube lens instead of 4-*f* system as in Figure S12, Supporting Information.

**Author contribution:** H.K. and J.J. contributed equally to this work. H.K. and J. J. conceived the idea, and J.S. supervised the project. J.J. conducted the theoretical analyses. H.K. performed the numerical simulations and designed the samples. H.K. and J.J. fabricated the samples and characterized them optically. H.K., J.J. and J.S. prepared the manuscript. All authors have accepted responsibility for the entire content of this submitted manuscript and approved its submission.

**Research funding:** This work is supported by National Research Foundation (NRF) grants (NRF-2021R1A2C2008687, NRF-2021M3H4A1A04086555) funded by the Ministry of Science and ICT (MSIT), Republic of Korea.

**Conflict of interest statement:** The authors declare no conflicts of interest.

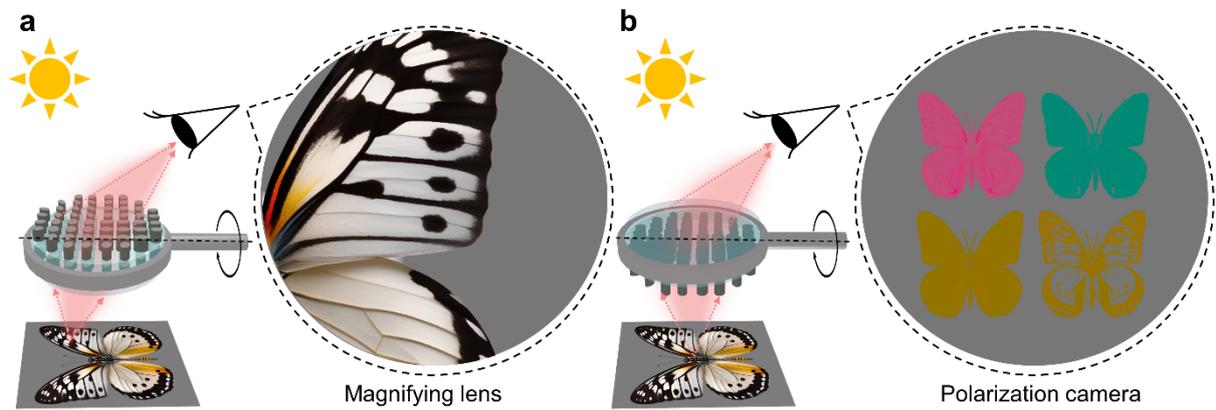

**Figure 1. Schematics of a device with asymmetric transmission.** (a) The device functions as a magnifying lens for the front-side illumination. (b) The device serves as a polarization camera for the back-side illumination. Note that the colors represent the polarization states of light, as described in a later section.

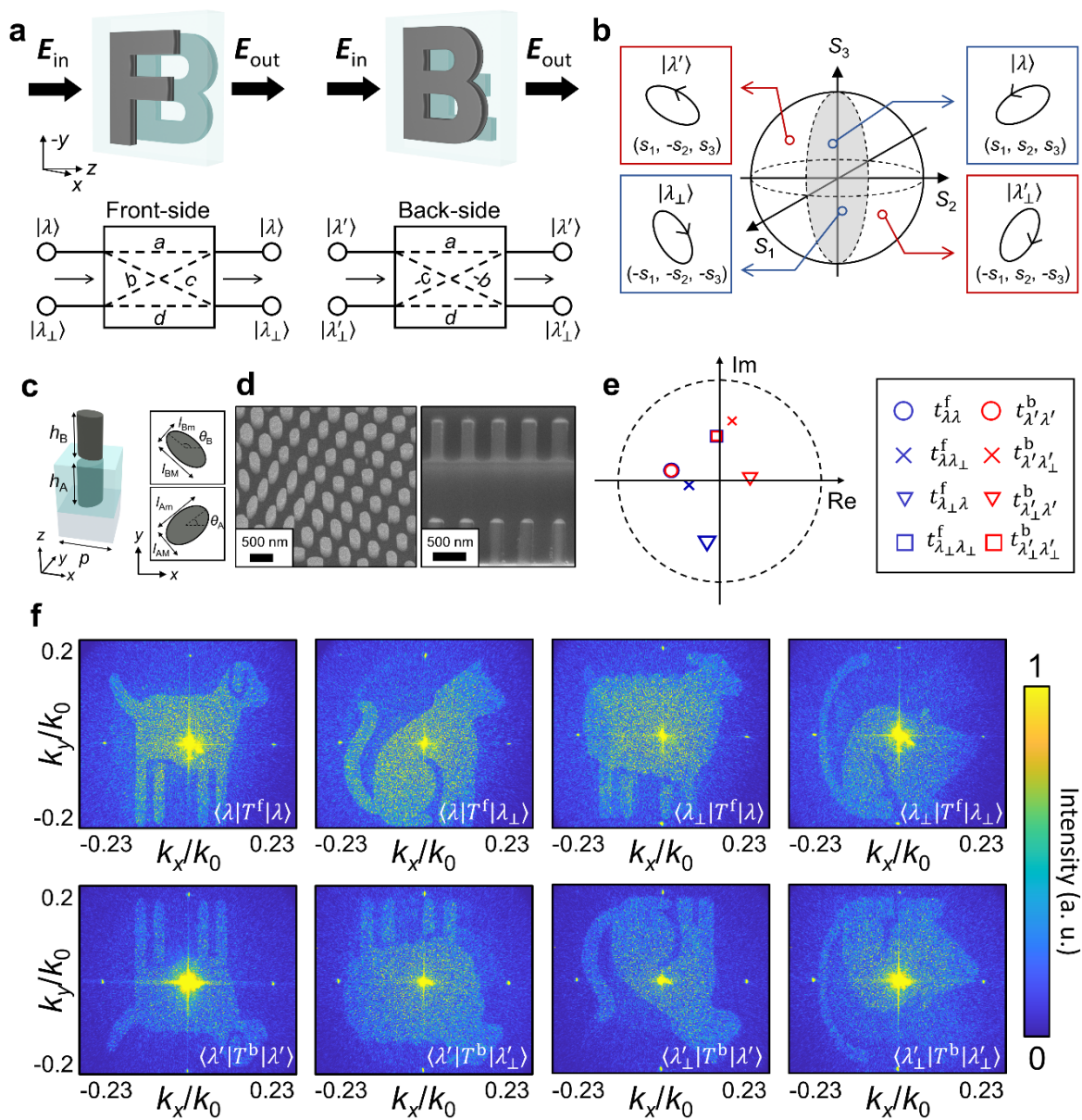

**Figure 2. Generalized asymmetric transmission.** (a) Schematics of optical system for the front- and back-side illuminations (upper panel), and the network representations of their Jones matrices (lower panel). The letters of "F" and "B" denote the front- and back-sides of the optical system, respectively. (b) Polarization bases for generalized asymmetric transmission on the Poincaré sphere. (c) Schematics of unit cell structure of bi-layer metasurfaces. The right panels show cross sections of the upper (the subscript B) and lower (the subscript A) layers of bi-layer structure. $p$: period; $h$: height of posts; $l_{M,m}$: length of the major (M) and minor (m) axes; $\theta$: orientation angle. (d) Scanning electron microscope images of a fabricated metasurface. (e) Numerical verification of generalized asymmetric transmission. The elements of the Jones matrices for the front- and back-side illuminations relative to the $\lambda\lambda_\perp$- and $\lambda'\lambda'_\perp$- polarization bases, respectively. (f) Experimental demonstration of generalized asymmetric transmission. $T^f$ and $T^b$ represent the Jones matrices for the front- and back-side illuminations, respectively. Kets represent the polarization of incident light. Bras represent the passing polarization of the polarizer.

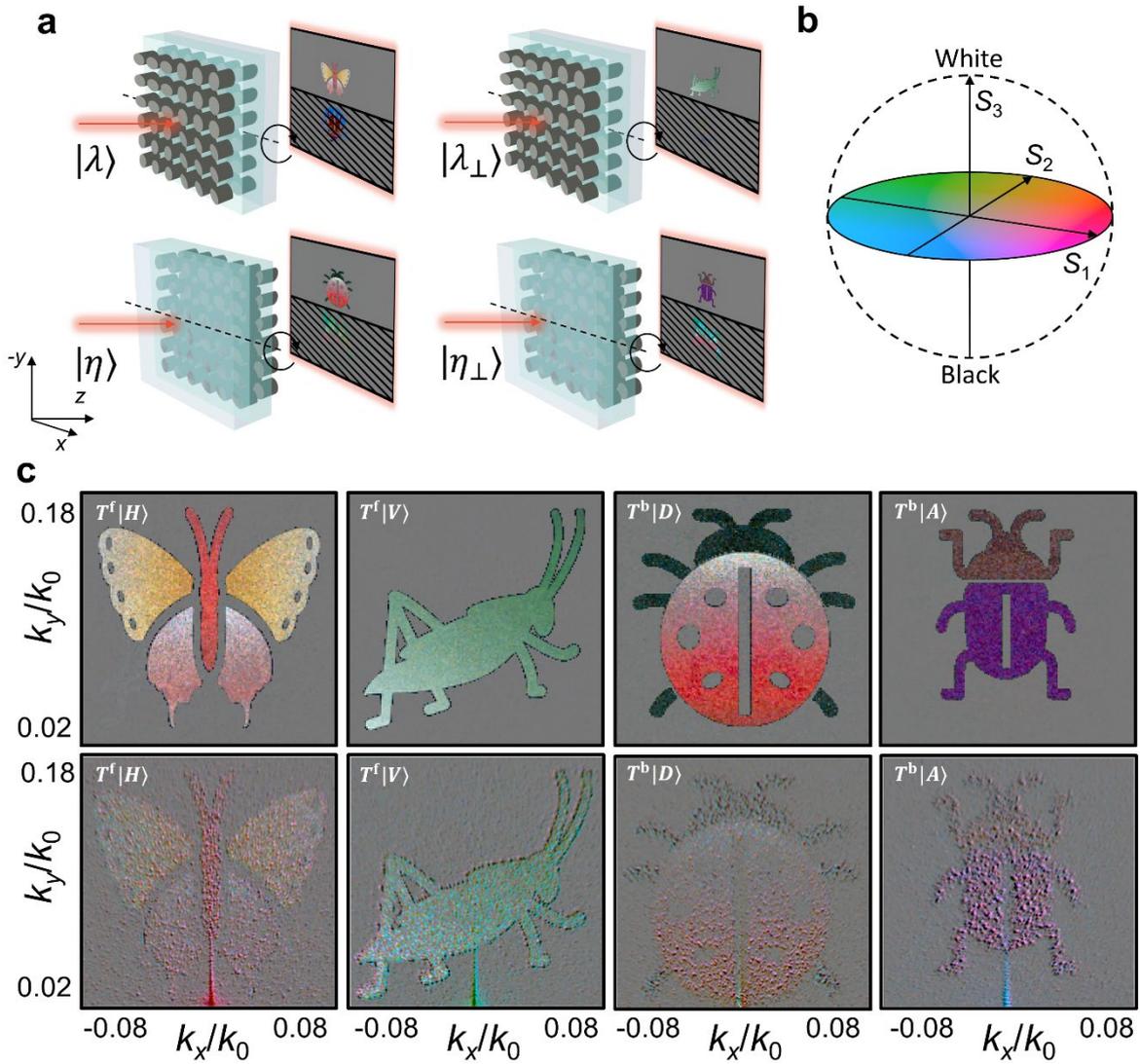

**Figure 3. Polarization-direction-multiplexed Janus vector holograms.** (a) Schematics of Janus vector holograms by a single bi-layer metasurface. Shaded regions indicate non-target spatial channels. (b) Mapping of CIELAB color space onto the normalized Poincaré sphere. The conversion relationship is given as $(L^*, a^*, b^*) = (50\cdot(S_3+1), 100\cdot S_1, 100\cdot S_2)$. (c) Experimental demonstration of Janus vector holograms. The upper row represents the target polarization images. The lower row represents the measured polarization images. $T^f$ and $T^b$ represent the Jones matrices for the front- and back-side illuminations, respectively. Kets represent the polarization of incident light; $|H\rangle$ for *x*-polarization; $|V\rangle$ for *y*-polarization; $|D\rangle$ for diagonal polarization; $|A\rangle$ for anti-diagonal polarization.

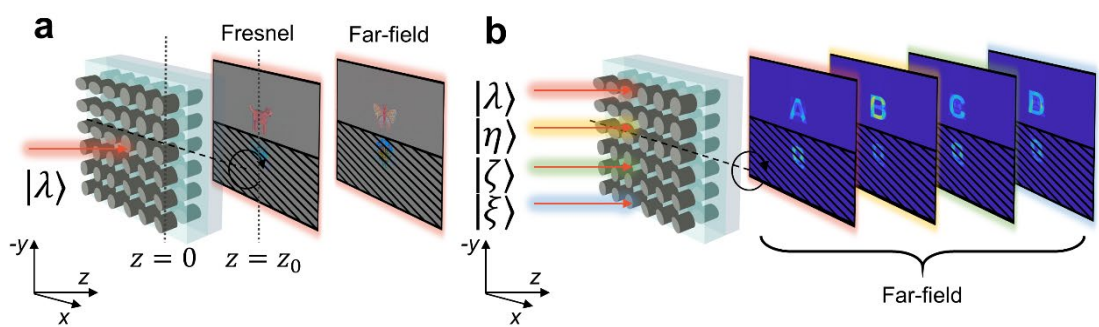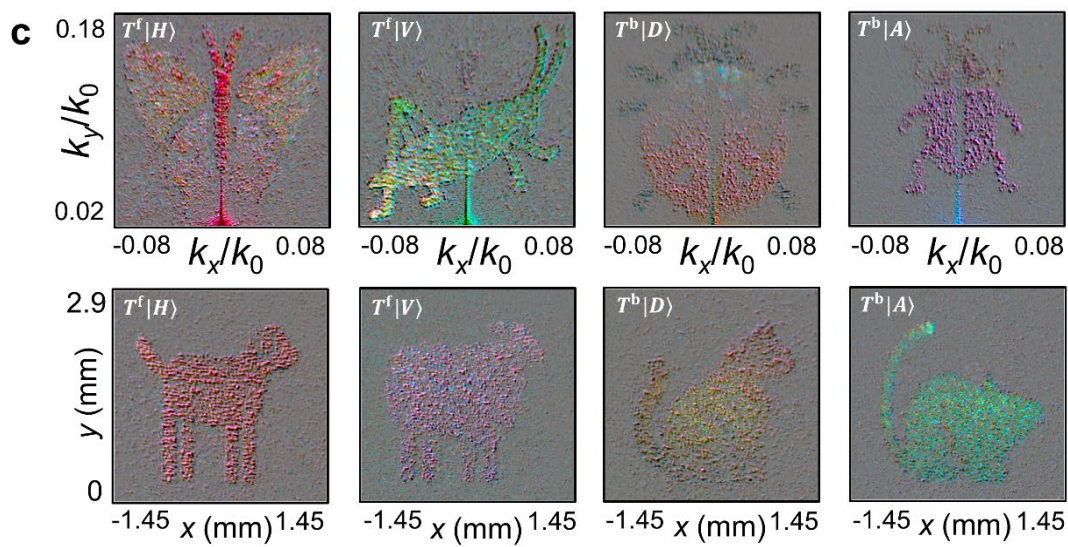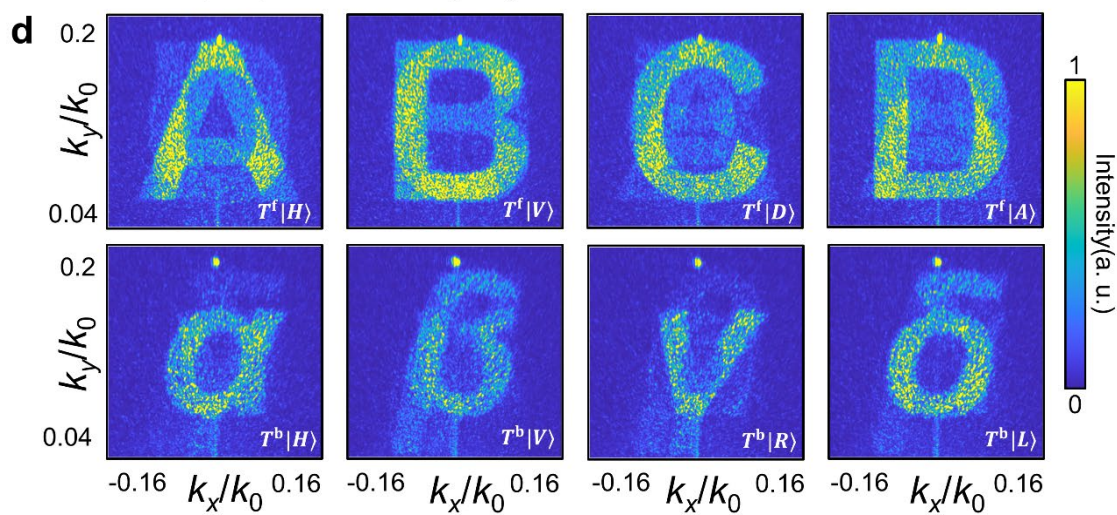

**Figure 4. Integration of Janus vector hologram and state-of-the-art hologram techniques.** (a) Schematics of multi-plane holograms. (b) Schematics of multichannel holograms with nonseparable polarization transformation. Shaded regions indicate non-target spatial channels. (c) Experimental demonstration of multi-plane holograms. The upper row represents the measured polarization images in the far-field regime. The lower row represents the measured polarization images in the Fresnel diffraction regime of $z_0$ = 1 mm. (d) Experimental demonstration of multichannel holograms with nonseparable polarization transformation. Each panel shows the measured intensity image with various incidence conditions. $T^f$ and $T^b$ represent the Jones matrices for the front- and back-side illuminations, respectively. Kets represent the polarization of incident light; $|H\rangle$ for x-polarization; $|V\rangle$ for y-polarization; $|D\rangle$ for diagonal polarization; $|A\rangle$ for anti-diagonal polarization; $|R\rangle$ for RCP polarization; $|L\rangle$ for LCP polarization.

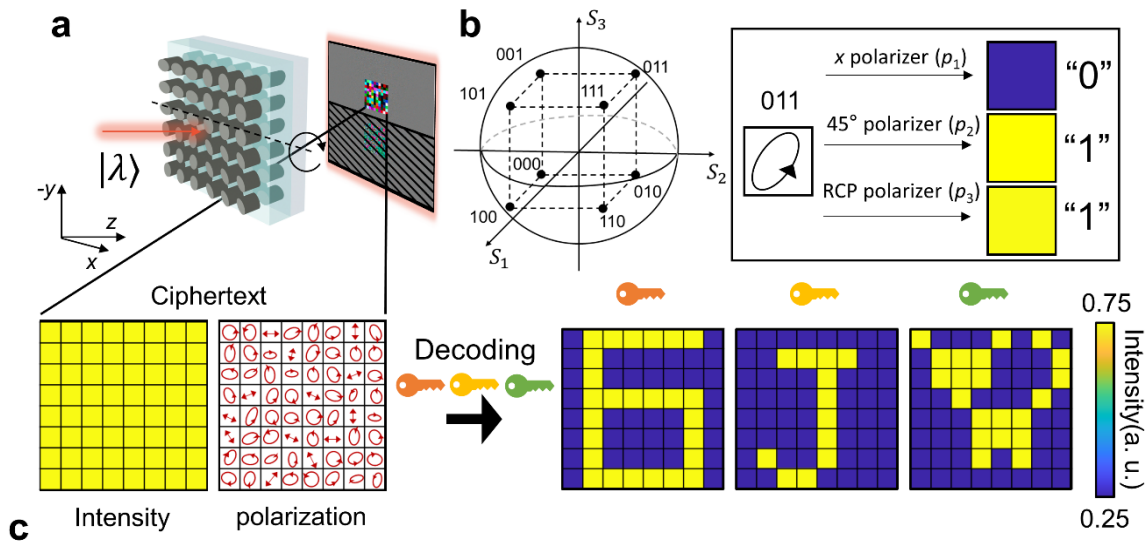

**Figure 5. Optical encryption based on Janus vector holograms.** (a) Schematics of optical encryption using a Janus metasurface generating a ciphertext with a pixelated vector polarization pattern. (b) Encryption and decoding mechanism involving polarization and polarizers. (c) Experimental demonstration of optical encryption. The left panels show the measured raw data of the Stokes parameters and polarization images. The right panels show the computationally decoded images with the appropriate VPs.

# Supporting Information for "Bidirectional Vectorial Holography Using Bi-Layer Metasurfaces and Its Application to Optical Encryption"

*Hyeonhee Kim*[1,†], *Joonkyo Jung*[1,†] and *Jonghwa Shin*[1,*]

[1]Department of Materials Science and Engineering, KAIST, Daejeon 34141, Republic of Korea.

[†]These authors equally contributed to this work.

[*]E-mail: qubit@kaist.ac.kr

**Supporting Texts**

**S1. Matrix analysis for the generalized asymmetric transmission**

Let's assume that the Jones matrix for the front-side illumination relative to the $x$- and $y$-polarization bases is given as

$$T^{\mathrm{f}}_{xy} = [|\lambda\rangle \quad |\lambda_\perp\rangle] \begin{bmatrix} a & b \\ c & d \end{bmatrix} \begin{bmatrix} \langle\lambda| \\ \langle\lambda_\perp| \end{bmatrix}.$$

As mentioned in the main text, if the given optical system is flipped around the $x$-axis, the Jones matrix for the back-side illumination relative to the $x$- and $y$-polarization bases in this case can be represented as

$$T^{\mathrm{b}}_{xy} = Q T^{\mathrm{r}}_{xy} Q = Q\left(T^{\mathrm{f}}_{xy}\right)^T Q = Q\left([|\lambda\rangle \quad |\lambda_\perp\rangle] \begin{bmatrix} a & b \\ c & d \end{bmatrix} \begin{bmatrix} \langle\lambda| \\ \langle\lambda_\perp| \end{bmatrix}\right)^T Q$$

$$= Q \begin{bmatrix} \langle\lambda| \\ \langle\lambda_\perp| \end{bmatrix}^T \begin{bmatrix} a & c \\ b & d \end{bmatrix} [|\lambda\rangle \quad |\lambda_\perp\rangle]^T Q$$

$$= Q \begin{bmatrix} \langle\lambda| \\ \langle\lambda_\perp| \end{bmatrix}^T Q \begin{bmatrix} a & -c \\ -b & d \end{bmatrix} Q [|\lambda\rangle \quad |\lambda_\perp\rangle]^T Q$$

$$= [|\lambda'\rangle \quad |\lambda'_\perp\rangle] \begin{bmatrix} a & -c \\ -b & d \end{bmatrix} \begin{bmatrix} \langle\lambda'| \\ \langle\lambda'_\perp| \end{bmatrix}$$

where $[|\lambda'\rangle \quad |\lambda'_\perp\rangle] = Q \begin{bmatrix} \langle\lambda| \\ \langle\lambda_\perp| \end{bmatrix}^T Q = \begin{bmatrix} 1 & 0 \\ 0 & -1 \end{bmatrix} \begin{bmatrix} \cos\chi & \sin\chi\,\mathrm{e}^{-\mathrm{i}\theta} \\ \sin\chi & -\cos\chi\,\mathrm{e}^{-\mathrm{i}\theta} \end{bmatrix}^T \begin{bmatrix} 1 & 0 \\ 0 & -1 \end{bmatrix} = \begin{bmatrix} \cos\chi & -\sin\chi \\ -\sin\chi\,\mathrm{e}^{-\mathrm{i}\theta} & -\cos\chi\,\mathrm{e}^{-\mathrm{i}\theta} \end{bmatrix}$, which corresponds to the Equation (3) in the main text.

For a general case where the axis of flipping is rotated by $\alpha$, the relation between the Jones matrices for the front- and back-side illuminations is given as $T^{\mathrm{b}}_{x'y'} = Q\left(T^{\mathrm{f}}_{x'y'}\right)^T Q$ where the subscript $x'y'$ represents the $x'$- and $y'$-polarization bases. These bases consist of the $x'$- and $y'$-axes that are rotated from the $x$- and $y$-axes by $\alpha$ as in Figure S2. Using the same derivation, if the Jones matrix for the front-side illumination relative to the $x'$- and $y'$-polarization bases is given as $T^{\mathrm{f}}_{x'y'} = [|\mu\rangle \quad |\mu_\perp\rangle] \begin{bmatrix} a & b \\ c & d \end{bmatrix} \begin{bmatrix} \langle\mu| \\ \langle\mu_\perp| \end{bmatrix}$ where $|\mu\rangle = [\cos\chi'\,;\sin\chi'\,\mathrm{e}^{\mathrm{i}\theta'}]$ and $|\mu_\perp\rangle = [\sin\chi'\,;-\cos\chi'\,\mathrm{e}^{\mathrm{i}\theta'}]$ representing polarization relative to the $x'$- and $y'$-polarization bases, the Jones matrix for the back-side illumination is given as $T^{\mathrm{b}}_{x'y'} = [|\mu'\rangle \quad |\mu'_\perp\rangle] \begin{bmatrix} a & -c \\ -b & d \end{bmatrix} \begin{bmatrix} \langle\mu'| \\ \langle\mu'_\perp| \end{bmatrix}$ where $|\mu'\rangle = [\cos\chi'\,;-\sin\chi'\,\mathrm{e}^{-\mathrm{i}\theta'}]$ and $|\mu'_\perp\rangle =$

$[-\sin\chi'; -\cos\chi' e^{-i\theta'}]$. Then, by transforming coordinates from the $x'$- and $y'$-polarization bases to the $x$- and $y$-polarization bases, the Jones matrices for the front- and back-side illuminations are given as

$$T_{xy}^{f} = R(\alpha)T_{x'y'}^{f}R(-\alpha) = R(\alpha)[|\mu\rangle \quad |\mu_{\perp}\rangle]\begin{bmatrix}a & b\\ c & d\end{bmatrix}\begin{bmatrix}\langle\mu|\\ \langle\mu_{\perp}|\end{bmatrix}R(-\alpha)$$

$$= [|\lambda\rangle \quad |\lambda_{\perp}\rangle]\begin{bmatrix}a & b\\ c & d\end{bmatrix}\begin{bmatrix}\langle\lambda|\\ \langle\lambda_{\perp}|\end{bmatrix},$$

$$T_{xy}^{b} = R(\alpha)T_{x'y'}^{b}R(-\alpha) = R(\alpha)[|\mu'\rangle \quad |\mu'_{\perp}\rangle]\begin{bmatrix}a & -c\\ -b & d\end{bmatrix}\begin{bmatrix}\langle\mu'|\\ \langle\mu'_{\perp}|\end{bmatrix}R(-\alpha)$$

$$= [|\lambda'\rangle \quad |\lambda'_{\perp}\rangle]\begin{bmatrix}a & -c\\ -b & d\end{bmatrix}\begin{bmatrix}\langle\lambda'|\\ \langle\lambda'_{\perp}|\end{bmatrix},$$

where the rotation matrix is $R(\alpha) = \begin{bmatrix}\cos\alpha & -\sin\alpha\\ \sin\alpha & \cos\alpha\end{bmatrix}$. From this expression, the polarization bases for the front- and back-side illuminations of the asymmetric transmission are given as

$$[|\lambda\rangle \quad |\lambda_{\perp}\rangle] = R(\alpha)[|\mu\rangle \quad |\mu_{\perp}\rangle] = R(\alpha)\begin{bmatrix}\cos\chi' & \sin\chi'\\ \sin\chi' e^{i\theta'} & -\cos\chi' e^{i\theta'}\end{bmatrix},$$

$$[|\lambda'\rangle \quad |\lambda'_{\perp}\rangle] = R(\alpha)[|\mu'\rangle \quad |\mu'_{\perp}\rangle] = R(\alpha)\begin{bmatrix}\cos\chi' & -\sin\chi'\\ -\sin\chi' e^{-i\theta'} & -\cos\chi' e^{-i\theta'}\end{bmatrix}.$$

Since $|\lambda\rangle$ and $|\lambda'\rangle$ are obtained from rotating $|\mu\rangle$ and $|\mu'\rangle$ by $\alpha$, their polarization ellipses can also be obtained by simply rotating the polarization ellipses, which are specified by $\chi'$ and $\theta'$, by $\alpha$. In Figure S2(c) and S2(d), the polarization ellipses and the angular positions on the Poincaré sphere of $|\lambda\rangle$ and $|\lambda'\rangle$ are described. From this geometrical visualization, it is clear that $|\lambda\rangle$ and $|\lambda'\rangle$ are mirror symmetric to each other with respect to the $S'_1$-$o$-$S_3$ plane as represented in Figure S2. Note that if the rotation angle $\alpha$ is zero, then the given polarization bases reduce to those in the main text.

## S2. Arbitrary Jones matrices achieved by universal metasurfaces

Universal metasurfaces are composed of two different types of bi-layer structures, as schematically illustrated in Figure S3. It is widely known that single-layer dielectric metasurfaces can be mathematically approximated as unitary symmetric Jones matrices ($U_S$) [1]. If we consider bi-layer structures constructed by cascading two dielectric metasurfaces, the Jones matrix of the bi-layer structures can be represented as a multiplication of two unitary symmetric matrices, which results in a unitary matrix ($U$) [2]. Furthermore, since two different types of bi-layer structures are interwoven to construct universal metasurfaces, the Jones matrix of the entire structure can be represented as an average of two unitary matrices. Given that by singular value decomposition, an arbitrary matrix ($A$) with singular values smaller than unity can be represented as an average of two unitary matrices, universal metasurfaces can generate arbitrary Jones matrices.

## S3. Gradient-descendent optimization

The required phase profiles for holograms in the main text were optimized using gradient-based optimization. Through the optimization process, the gradient of the given loss function with respect to the design parameters can be obtained using the chain rule. Then, the design parameters can be updated in the direction of reducing the loss function based on the obtained gradient. As iterations proceed, the optimal design parameters that generate desired set of holograms as closely as possible can be obtained by minimizing the loss function. The flowchart describing the entire optimization process is given in Figure S11.

If any loss function is given as a function of far-field components, the gradient can be analytically related to the local design parameters as follows [3]. First, the derivatives of the loss function with respect to far-field components, $\frac{\partial L}{\partial E^f_{pq,x}}$ and $\frac{\partial L}{\partial E^f_{pq,y}}$ where $L$ is the loss function, $E^f$ is the electric field in the far-field regime, the subscript $p$ and $q$ represent indices of the spatial frequency domain and the subscript $x$ and $y$ represent the $x$- and $y$-polarization components of the field, can be readily obtained from the definition of the loss functions. Then, by the chain rule, the derivatives of the loss function with respect to the electric field right after the optical system is given as

$$\frac{\partial L}{\partial E^n_{ab,(x,y)}} = \sum_{p,q} \frac{\partial L}{\partial E^f_{pq,(x,y)}} \frac{\partial E^f_{pq,(x,y)}}{\partial E^n_{ab,(x,y)}}$$

where $E^n$ is the electric field right after the optical system and the subscript $a$ and $b$ represent the indices of the spatial domain. Note that by the discrete Fourier transform, which is given as $E^f_{pq,(x,y)} = \mathcal{F}(E^n_{(x,y)})_{pq} = \sum_{a,b} E^n_{ab,(x,y)} e^{-i(k_x^{(p)} x^{(a)} + k_y^{(q)} y^{(b)})}$ where $\mathcal{F}$ represents the Fourier transform and the superscripts $a$ and $b$ ($p$ and $q$) also represent indices of the spatial(spatial frequency) domain, the derivatives of the far-field components are given as $\frac{\partial E^f_{pq,(x,y)}}{\partial E^n_{ab,(x,y)}} = e^{-i(k_x^{(p)} x^{(a)} + k_y^{(q)} y^{(b)})}$. With this, the above derivatives of the loss function with respect to the electric field right after the optical system can be re-written as

$$\frac{\partial L}{\partial E^n_{ab,(x,y)}} = \sum_{p,q} \frac{\partial L}{\partial E^f_{pq,(x,y)}} e^{-i(k_x^{(p)} x^{(a)} + k_y^{(q)} y^{(b)})} = N^2 \left[ \mathcal{F}^{-1} \left\{ \left( \frac{\partial L}{\partial E^f_{(x,y)}} \right)^* \right\} \right]_{ab}^*$$

where $N$ is the number of points in the spatial domain and the superscript * represents the complex conjugate. Furthermore, since this electric field right after the optical system is determined from the incident electric field and the Jones matrix of the optical system, we can obtain the derivatives of the loss function with respect to the design parameters again by the chain rule. To maximally utilize the incident energy, we assumed our metasurfaces have unitary Jones matrices; in this case, our design parameters was set to be $\phi_1$, $\phi_2$, $\theta_1$ and $\theta_2'$, which can generate arbitrary unitary matrices given as $U = \begin{bmatrix} \cos\theta_2' & -\sin\theta_2' \\ \sin\theta_2' & \cos\theta_2' \end{bmatrix} \begin{bmatrix} e^{i\phi_1} & 0 \\ 0 & e^{i(\phi_2+\pi)} \end{bmatrix} \begin{bmatrix} \cos\theta_1 & \sin\theta_1 \\ -\sin\theta_1 & \cos\theta_1 \end{bmatrix} = \begin{bmatrix} U_{11} & U_{12} \\ U_{21} & U_{22} \end{bmatrix}$. These design parameters are directly related to the structural parameters of metasurfaces as explained in [2]. Then, if we assume plane wave incidence, the electric field right after the metasurface can be given as $\begin{bmatrix} E^n_{ab,x} \\ E^n_{ab,y} \end{bmatrix} = \begin{bmatrix} U_{11} & U_{12} \\ U_{21} & U_{22} \end{bmatrix} \begin{bmatrix} E^{in}_x \\ E^{in}_y \end{bmatrix} = \begin{bmatrix} U_{11} E^{in}_x + U_{12} E^{in}_y \\ U_{21} E^{in}_x + U_{22} E^{in}_y \end{bmatrix}$ where $E^{in}$ is the incident electric field. With this relation, the derivatives of the loss function with respect to the design parameters are given as

$$\frac{\partial L}{\partial \phi_{1,ab}} = \frac{\partial L}{\partial E^n_{ab,(x,y)}} \frac{\partial E^n_{ab,(x,y)}}{\partial \phi_{1,ab}} + \frac{\partial L}{\partial E^{n,*}_{ab,(x,y)}} \frac{\partial E^{n,*}_{ab,(x,y)}}{\partial \phi_{1,ab}}$$

$$\frac{\partial L}{\partial \phi_{2,ab}} = \frac{\partial L}{\partial E^n_{ab,(x,y)}} \frac{\partial E^n_{ab,(x,y)}}{\partial \phi_{1,ab}} + \frac{\partial L}{\partial E^{n,*}_{ab,(x,y)}} \frac{\partial E^{n,*}_{ab,(x,y)}}{\partial \phi_{2,ab}}$$

$$\frac{\partial L}{\partial \theta_{1,ab}} = \frac{\partial L}{\partial E^n_{ab,(x,y)}} \frac{\partial E^n_{ab,(x,y)}}{\partial \theta_{1,ab}} + \frac{\partial L}{\partial E^{n,*}_{ab,(x,y)}} \frac{\partial E^{n,*}_{ab,(x,y)}}{\partial \theta_{1,ab}}$$

$$\frac{\partial L}{\partial \theta'_{2,ab}} = \frac{\partial L}{\partial E^n_{ab,(x,y)}} \frac{\partial E^n_{ab,(x,y)}}{\partial \theta'_{2,ab}} + \frac{\partial L}{\partial E^{n,*}_{ab,(x,y)}} \frac{\partial E^{n,*}_{ab,(x,y)}}{\partial \theta'_{2,ab}}.$$

For holograms generating images within the Fresnel diffraction regime, the electric field in this regime should be considered. In this case, a loss function is given as a function of the electric field within the Fresnel diffraction regime. The derivatives of the loss function with respect to the electric field in the Fresnel diffraction regime, $\frac{\partial L}{\partial E^{n'}_{a'b',x}}$ and $\frac{\partial L}{\partial E^{n'}_{a'b',y}}$ where $E^{n'}$ is the electric field at the plane of interest distant from the optical system and the subscripts $a'$ and $b'$ represent indices of the spatial domain, can be readily obtained from the definition. By considering free-space propagation using the Fourier transform, the derivatives of the given loss function can also be analytically related to the local design parameters as follows. The electric field after propagating the distance, $d$, from the optical system is given as $E^{n'}_{a'b',(x,y)} = N^{-2} \sum_{pq} E^f_{pq,(x,y)} e^{ik_z^{(p,q)} d} e^{i\left(k_x^{(p)} x^{(a)} + k_y^{(q)} y^{(b)}\right)} = \mathcal{F}^{-1}\left(E^f_{(x,y)} e^{ik_z d}\right)_{a'b'}$. Then, the derivatives of the loss function with respect to the far-field components is given as

$$\frac{\partial L}{\partial E^f_{pq,(x,y)}} = \sum_{a',b'} \frac{\partial L}{\partial E^{n'}_{a'b',(x,y)}} \frac{\partial E^{n'}_{a'b',(x,y)}}{\partial E^f_{pq,(x,y)}} = N^{-2} e^{ik_z^{(p,q)} d} \left[\mathcal{F}\left\{\left(\frac{\partial L}{\partial E^{n'}_{(x,y)}}\right)^*\right\}_{pq}\right]^*.$$

Then, the derivatives of the loss function with respect to the electric field right after the optical system is given as

$$\frac{\partial L}{\partial E^n_{ab,(x,y)}} = \sum_{p,q} \frac{\partial L}{\partial E^f_{pq,(x,y)}} \frac{\partial E^f_{pq,(x,y)}}{\partial E^n_{ab,(x,y)}} = N^2 \left[\mathcal{F}^{-1}\left\{\left(\frac{\partial L}{\partial E^f_{(x,y)}}\right)^*\right\}_{ab}\right]^*.$$

Finally, the derivatives of the loss function with respect to the design parameters can be represented as the same form with the previous case.

# Supporting Figures

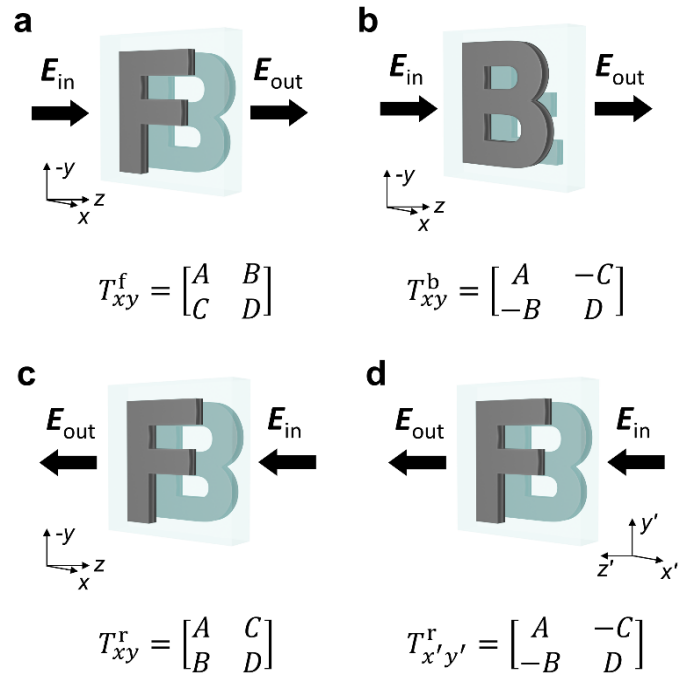

**Figure S1. Equivalence of flipping an optical system and flipping the direction of illumination.** (a) Original geometry of the optical system and incident light. (b) Geometry of the flipped optical system. (c) Geometry of the reciprocal scenario where the incident direction is flipped. (d) Geometry of the reciprocal scenario with coordinate transformation. Note that the Jones matrix has the same form, but the basis is different for S1(b) and S1(d).

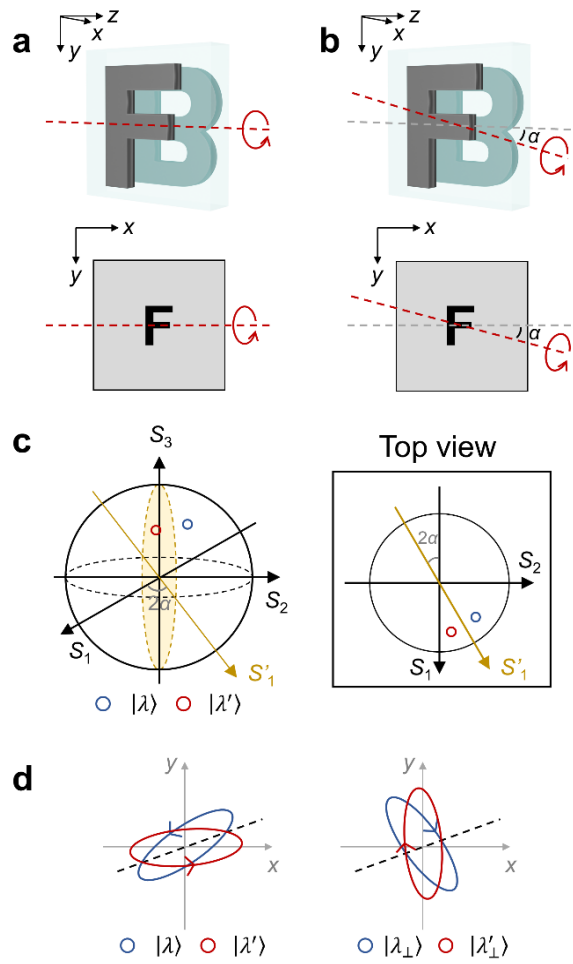

**Figure S2. Rotation of flipping axis.** (a) Geometry of an optical system flipped around x-axis with a rotation angle of $\alpha = 0°$. (b) Geometry of an optical system flipped around an axis with arbitrary rotation angle of $\alpha$. (c) Relation between polarizations of $|\lambda\rangle$ and $|\lambda'\rangle$ on the Poincaré sphere. (d) Polarization ellipses of the two polarization bases.

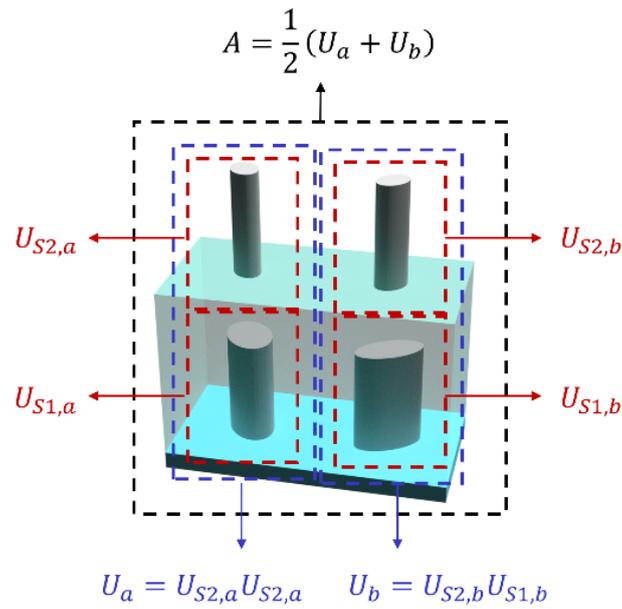

**Figure S3. Matrix-based analysis of universal metasurfaces.** Each post can be approximately represented as a unitary symmetric Jones matrix $U_S$ (red dotted box). The Jones matrix of each cluster in bi-layer structure is given as a multiplication of the Jones matrices of the lower layer (noted by the subscript 1) and upper layer (noted by the subscript 2), resulting in a unitary Jones matrix $U$ (blue dotted box). The Jones matrix of the supercell structure with two types of clusters (noted by the subscripts *a* and *b*) can be represented as an average of two unitary matrices, resulting in an arbitrary Jones matrix $A$ (black dotted box).

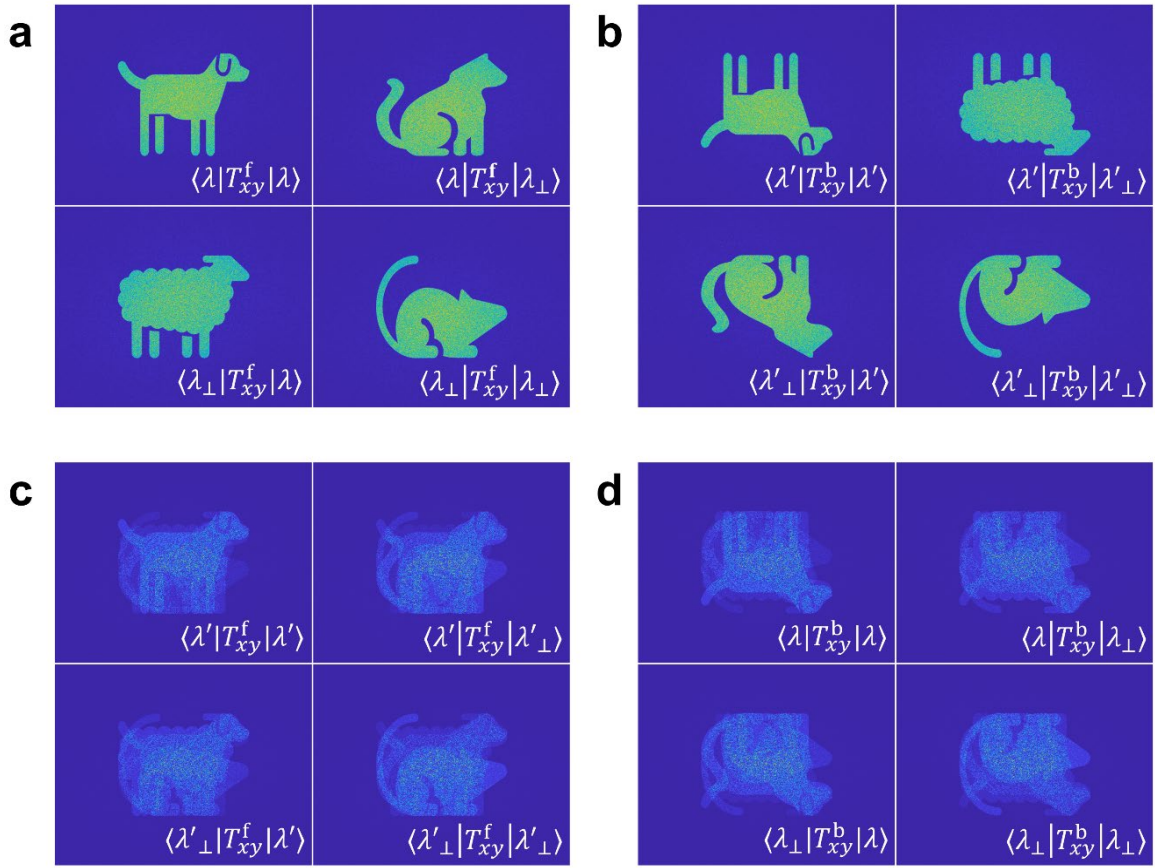

**Figure S4. Extended data of scalar holograms for the generalized asymmetric transmission.** (a) Optimized intensity patterns for the front-side illumination with $\lambda$- and $\lambda_\perp$-polarizations. (b) Optimized intensity patterns for the back-side illumination with $\lambda'$- and $\lambda'_\perp$-polarizations. (c) Numerically obtained intensity patterns for the front-side illumination with $\lambda'$- and $\lambda'_\perp$-polarizations. (d) Numerically obtained intensity patterns for the back-side illumination with $\lambda$- and $\lambda_\perp$-polarizations.

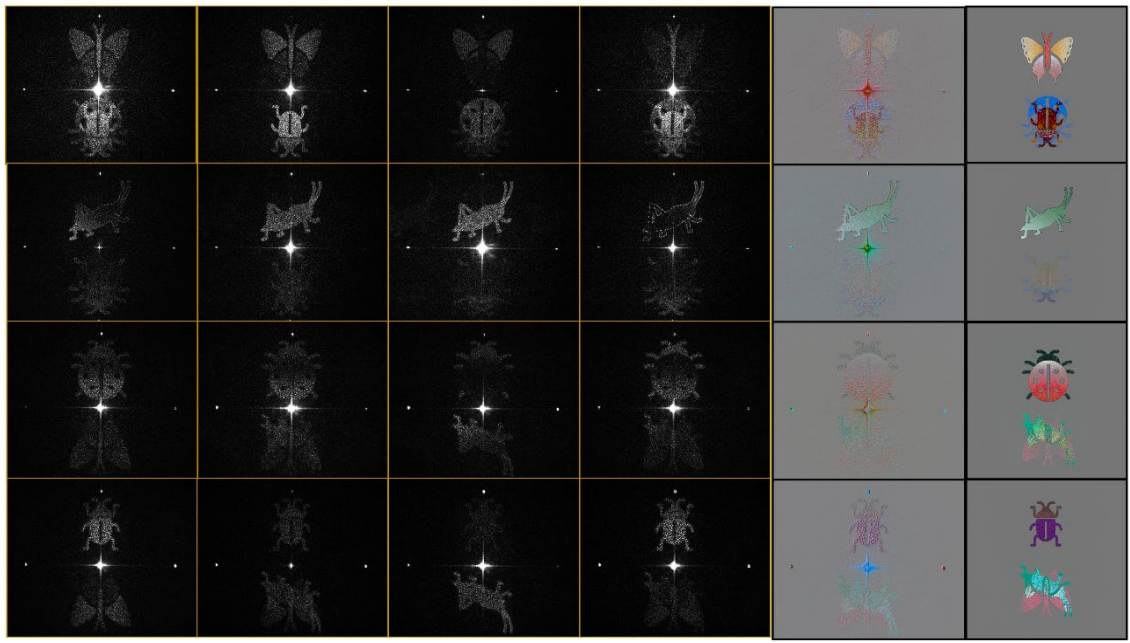

**Figure S5. Extended data of polarization-direction-multiplexed Janus vector holograms.** The left panels (from the first to fourth columns) show raw data of the full-Stokes polarimetry. The right panels (from the fifth to sixth columns) show measured and optimized polarization patterns in full transmission space represented using the false color of CIELAB.

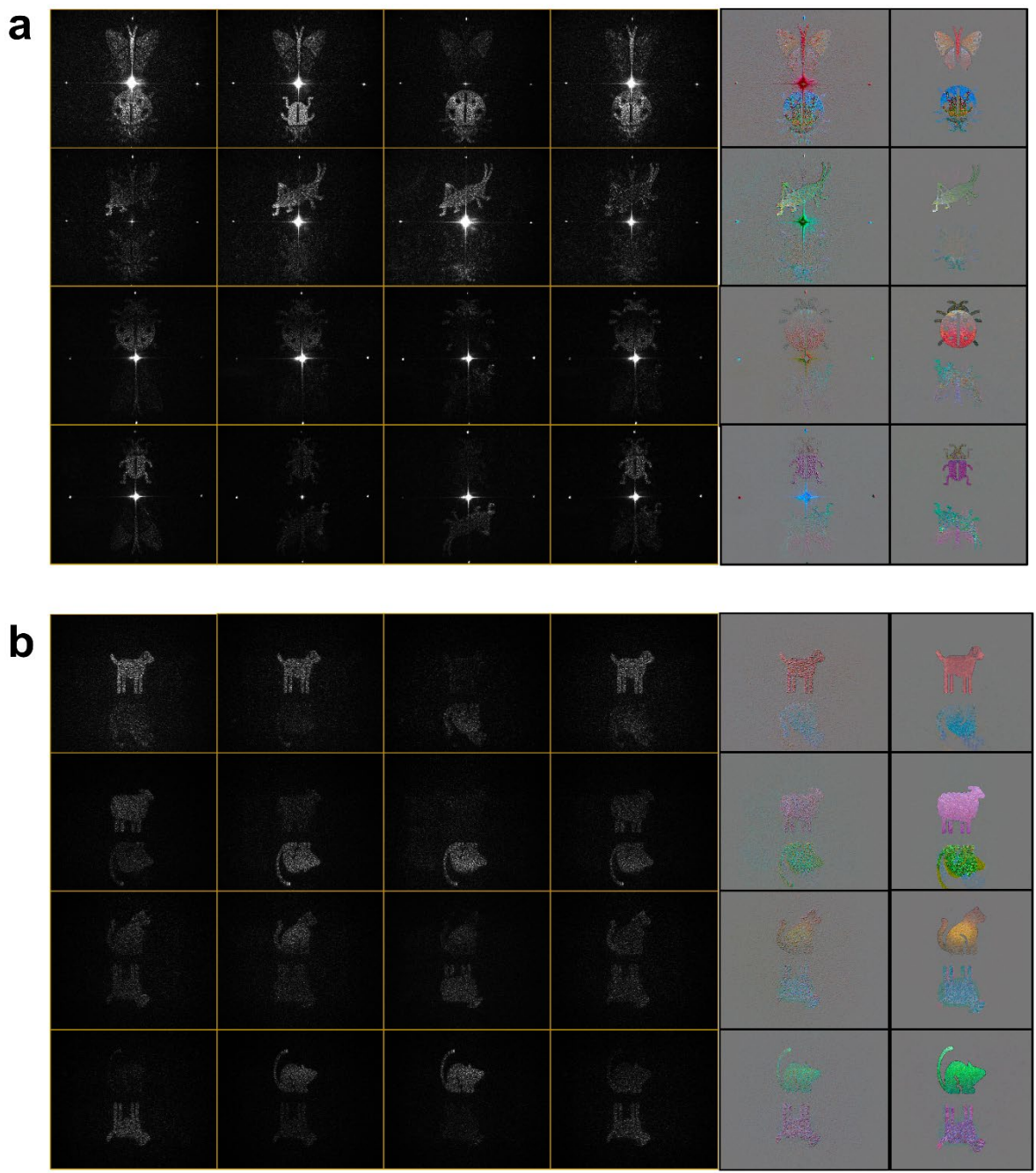

**Figure S6. Extended data of multi-plane vector holograms.** (a) Measured data in the far-field regime. (b) Measured data in the Fresnel diffraction regime. The left panels (from the first to fourth columns) show raw data of the full-Stokes polarimetry. The right panels (from the fifth to sixth columns) show measured and optimized polarization patterns in full transmission space represented using the false color of CIELAB.

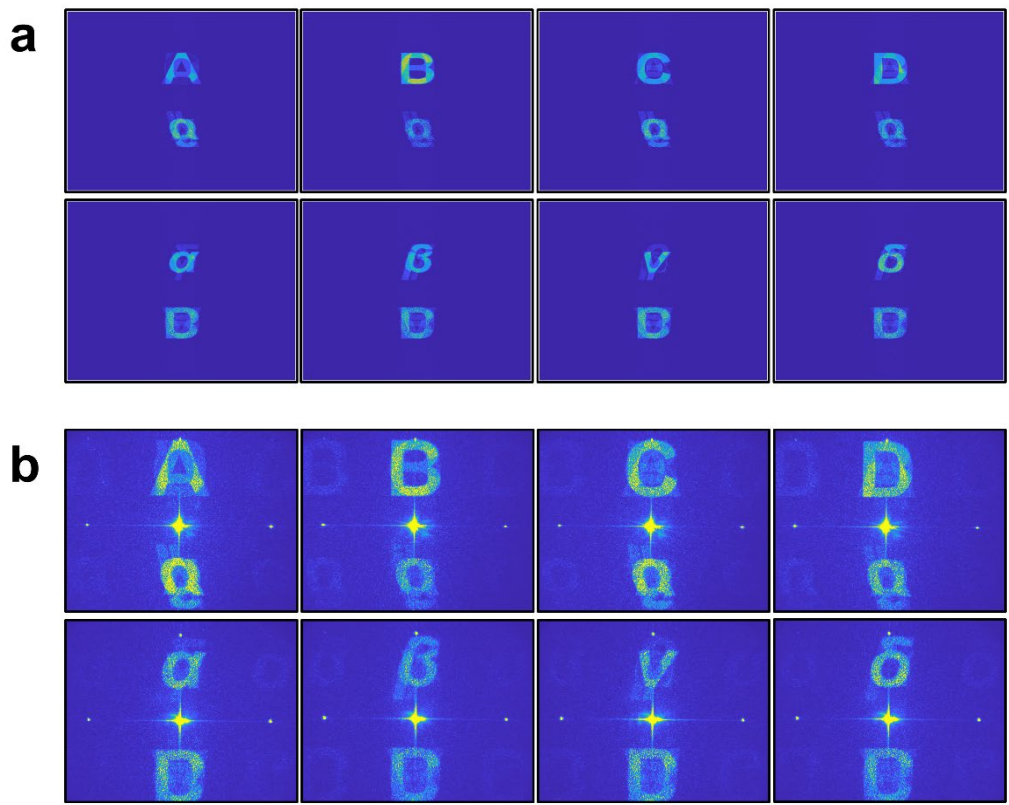

**Figure S7. Extended data of multichannel holograms with nonseparable polarization transformation.** (a) Optimized intensity images. (b) Measured intensity images in full transmission space.

**Figure S8.** Experimental demonstration of optical encryption with 8-by-8 pixelated polarization patterns.

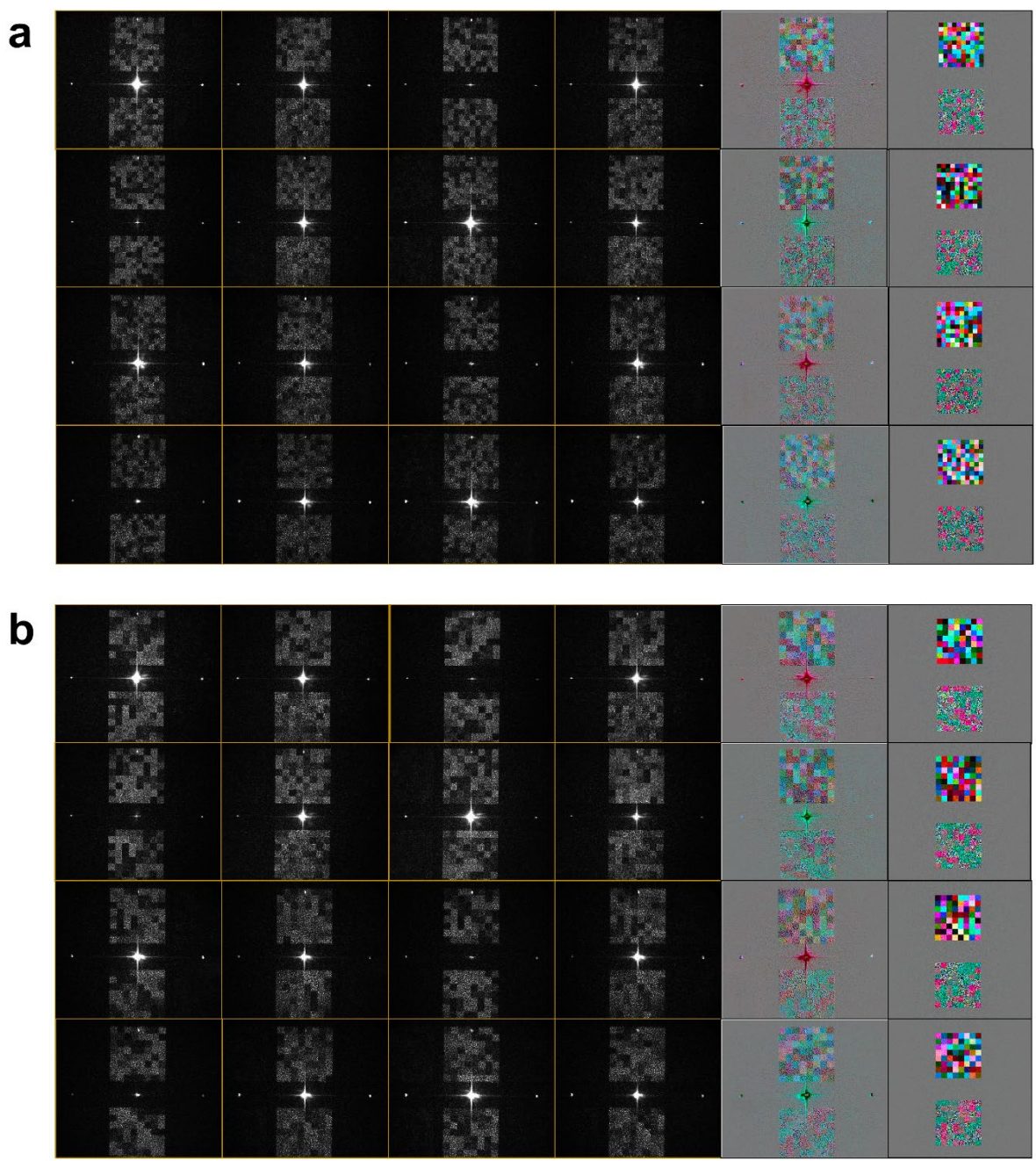

**Figure S9. Extended data of vector holograms for optical encryption.** (a) Optical encryption with 10-by-10 pixelated images. (b) Optical encryption with 8-by-8 pixelated images. The left panels (from the first to fourth columns) show raw data of the full-Stokes polarimetry. The right panels (from the fifth to sixth columns) show measured and optimized polarization patterns in full transmission space represented using the false color of CIELAB.

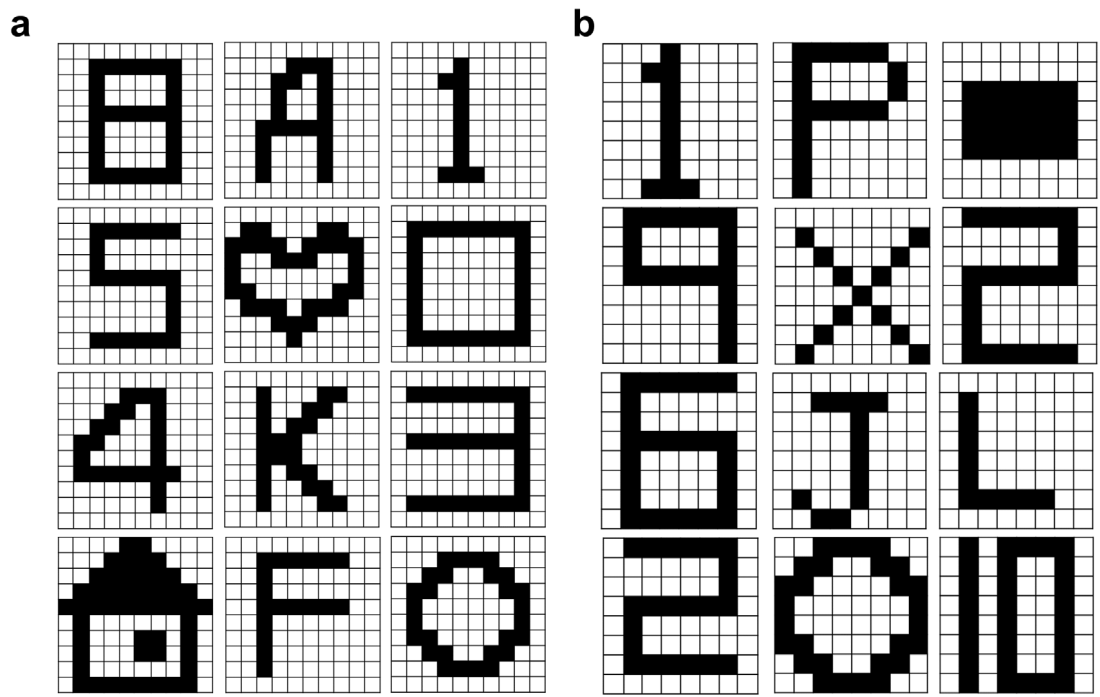

**Figure S10. Ground truth hidden information.** (a) 10-by-10 pixelated binary images. (b) 8-by-8 pixelated binary images.

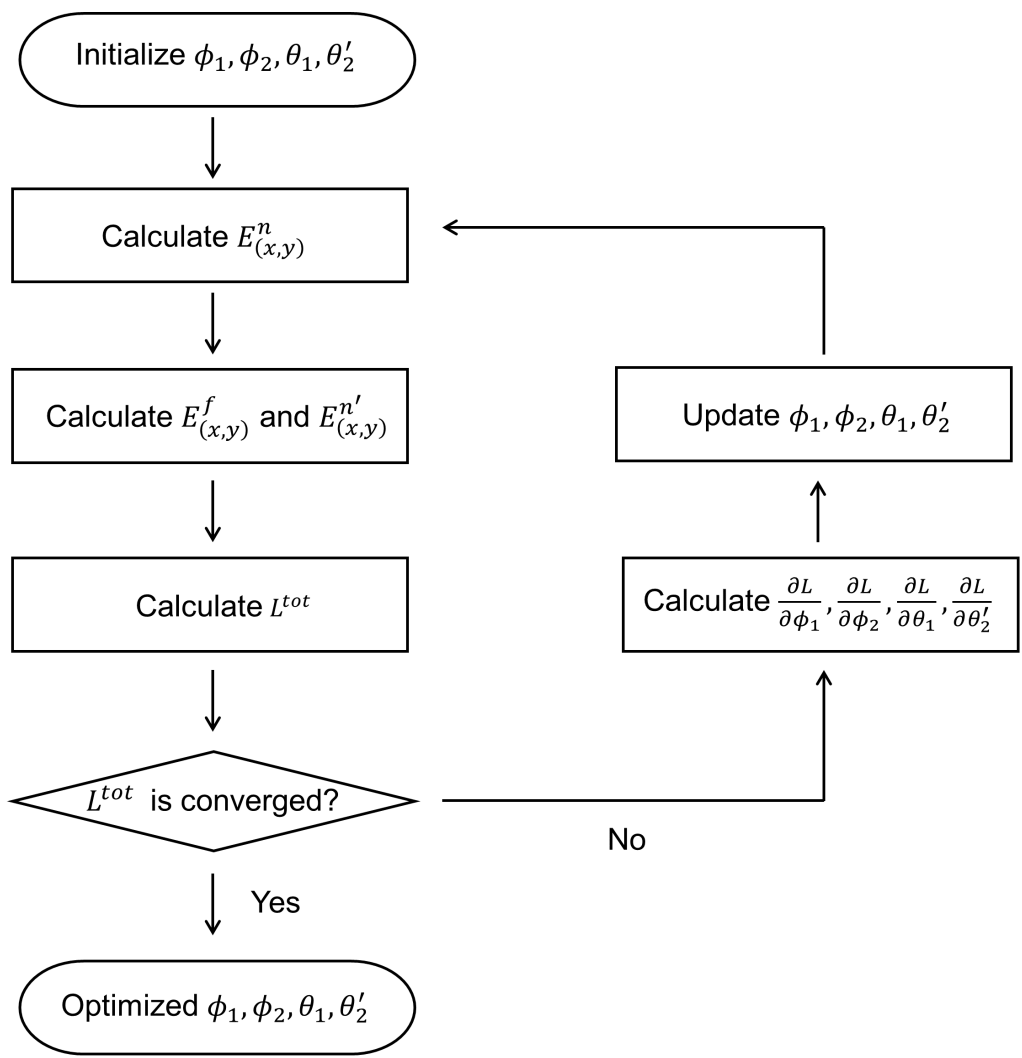

**Figure S11. Flowchart for gradient-descent optimization.**

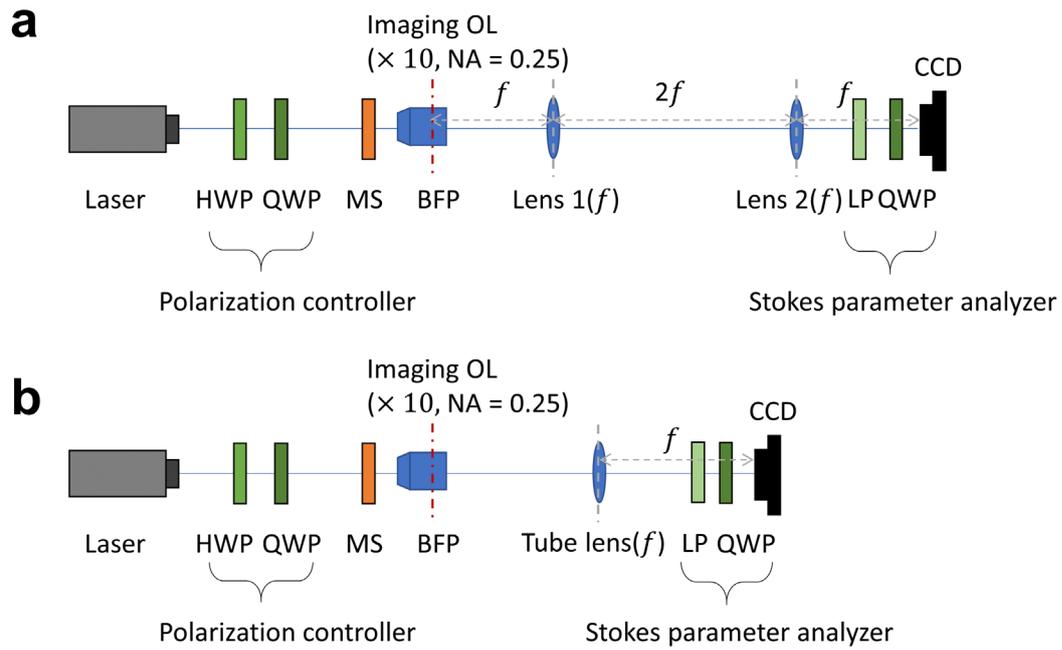

**Figure S12. Optical characterization.** (a) Setup for measuring holographic images in the far-field regime. (b) Setup for measuring holographic images in the Fresnel diffraction regime. HWP: half waveplate; QWP: quarter waveplate; MS: metasurfaces; OL: objective lens; BFP: back focal plane; $f$: distance amount of focal length; CCD: charge-coupled device; LP: linear polarizer.